\documentclass{aa}  
\usepackage{graphicx}
\usepackage{txfonts}
\usepackage{chemformula}
\let\ce\ch

\usepackage{orcidlink}
\newcommand{\orcid}[1]{\unskip\protect\href{https://orcid.org/#1}{\protect\includegraphics[width=8pt,clip]{logo_orcid}}}
\usepackage{xspace}
\usepackage{booktabs}
\usepackage{color}
\usepackage{amssymb}
\usepackage{pifont}
\newcommand{\cmark}{\ding{51}}%
\newcommand{\xmark}{\ding{55}}%

\usepackage{longtable}

\usepackage{cleveref}
\usepackage{newtxtext}
\usepackage[varvw]{newtxmath}

\begin{document} 

   \title{PRODIGE - Envelope to Disk with NOEMA\thanks{Based on observations carried out under project number L19MB with the IRAM NOEMA Interferometer. IRAM is supported by INSU/CNRS (France), MPG (Germany) and IGN (Spain)}}
    
   \subtitle{III. The origin of complex organic molecule emission in SVS13A}

   \author{T.-H. Hsieh
          \inst{1}
          \and          
          J. E. Pineda
          \inst{1}\orcidlink{0000-0002-3972-1978}
          \and
            D. M. Segura-Cox\thanks{NSF Astronomy and Astrophysics Postdoctoral Fellow}
          \inst{2,1}\orcidlink{0000-0003-3172-6763}
          \and
          P. Caselli
          \inst{1}\orcidlink{0000-0003-1481-7911}
          \and
          M. T. Valdivia-Mena
          \inst{1}\orcidlink{0000-0002-0347-3837}
          \and
          C. Gieser
          \inst{1}
          \and
          M. J. Maureira
          \inst{1}\orcidlink{0000-0002-7026-8163}
          \and
          A. Lopez-Sepulcre
          \inst{3,4}
          \and
          L. Bouscasse
          \inst{3}
          \and
          R. Neri
          \inst{3}\orcidlink{0000-0002-7176-4046}
          \and
          Th. M\"{o}ller
          \inst{5}
          \and
          A. Dutrey
          \inst{6}
          \and
          A. Fuente
          \inst{7}
          \and
          D. Semenov
          \inst{8}
          \and
          E. Chapillon
          \inst{3}
          \and
          N. Cunningham
          \inst{4}\orcidlink{0000-0003-3152-8564}
          \and
          Th. Henning
          \inst{8}
          \and
          V. Pietu
          \inst{3}
          \and
          I. Jimenez-Serra
          \inst{7}
            \and
          S. Marino
            \inst{9}
          \and
          C. Ceccarelli
          \inst{4}
          }

   \institute{Max-Planck-Institut f\"{u}r extraterrestrische Physik, Giessenbachstrasse 1, D-85748 Garching, Germany\\
              \email{thhsieh@mpe.mpg.de}
         \and
            Department of Astronomy, The University of Texas at Austin, 2500 Speedway, Austin, TX, 78712, USA
        \and
            Institut de Radioastronomie Millim\'{e}trique (IRAM), 300 rue de la Piscine, F-38406, Saint-Martin d'H\`{e}res, France
        \and
            IPAG, Universit\'{e} Grenoble Alpes, CNRS, F-38000 Grenoble, France
        \and
            I. Physikalisches Institut, Universität zu K\"{o}ln, Z\"{u}lpicher Str. 77, 50937 K\"{o}ln, Germany
        \and
            Laboratoire d'Astrophysique de Bordeaux, Universit\'{e} de Bordeaux, CNRS, B18N, All\'{e}e Geoffroy Saint-Hilaire, F-33615 Pessac, France
        \and
            Centro de Astrobiolog\'{\i}a (CAB), CSIC-INTA, Ctra.deTorrej\'on a Ajalvir km 4, 28806, Torrej\'on de Ardoz, Spain
        \and
            Max-Planck-Institut f\"{u}r Astronomie, K\"{o}nigstuhl 17, D-69117 Heidelberg, Germany
        \and 
            Department of Physics and Astronomy, University of Exeter, Stocker Road, Exeter, EX4 4QL, UK
             }

   \date{Mar 18, 2024}

 
  \abstract
   {Complex Organic Molecules (COMs) have been found toward low-mass protostars but the origins of the COM emission are still unclear. It can be associated with, for example, hot corinos, outflows, and/or accretion shock/disk atmosphere.}
   {We aim to disentangle the origin of the COM emission toward the chemically-rich protobinary system SVS13A using six O-bearing COMs.}
   {We have conducted NOrthern Extended Millimeter Array (NOEMA) observations toward SVS13A from the PROtostars \& DIsks: Global Evolution (PRODIGE) program. Our previous \ce{DCN} observations reveal a possible infalling streamer, which may affect the chemistry of the central protobinary by inducing accretion outbursts and/or shocked gas. Here, we further analyze six O-bearing COMs: \ce{CH3OH}, aGg'-\ce{(CH2OH)2}, \ce{C2H5OH}, \ce{CH2(OH)CHO}, \ce{CH3CHO}, and \ce{CH3OCHO}. Although the COM emission is not spatially resolved, we constrain the source sizes to $\lesssim0.3-0.4$ arcsec (90$-$120 au) by conducting uv-domain Gaussian fitting. Interestingly, the high-spectral resolution data reveal complex line profiles with multiple peaks; although the line emission is likely dominated by the secondary VLA4A at $V_{\rm LSR}=7.36$ km s$^{-1}$, the numbers of peaks ($\sim$$2-5$), velocities, and linewidths are different between these six O-bearing COMs. The LTE fitting unveils differences in excitation temperatures and emitting areas among these COMs.
   We further conduct multiple-velocity-component LTE fitting to decompose the line emission into different kinematic components. 
   As a result, the emission of these COMs is decomposed into up to 6 velocity components from the LTE modeling. 
   The physical conditions (temperature, column density, and source size) of these components from each COM are obtained, and (Markov chain Monte Carlo) MCMC sampling are performed to test the fitting results.}
   {We find a variety in excitation temperatures ($100-500$ K) and source sizes (D$\sim10-70$ au) from these kinematic components from different COMs. The emission of each COM can trace several components and different COMs most likely trace different regions.}
   {Given this complex structure, we suggest that the central region is inhomogeneous and unlikely to be heated by only protostellar radiation. 
   We conclude that accretion shocks induced by the large-scale infalling streamer likely exist and contribute to the complexity of the COM emission.
   This underlines the importance of high-spectral resolution data when analyzing COM emission in protostars and deriving relative COM abundances.}


   \maketitle
%
\section{Introduction} \label{sec:intro}
About 270 molecules are found in the interstellar medium (ISM) at present time \citep{ce22}. Among these molecules, 40\% are so-called interstellar Complex Organic Molecules (COMs), defined as carbon-bearing molecules that have at least 6 atoms \citep{he09,ce17}.
COMs are commonly detected in massive star-forming regions \citep{bl87,gi23} and are also occasionally detected toward solar-type protostellar systems (\citealt{ca12,va14}, and references therein).
In Class 0 protostars, COMs are seen in several cases such as IRAS16293-2422 \citep{va95,ca03,jo18}, NGC1333-IRAS4A \citep{ca03}, NGC1333-IRAS4B/IRAS2A \citep{bo07}.
COMs have even been detected toward prestellar cores (e.g. L1689B: \citealt{ba12}, L1544: \citealt{ta06,sp16}).
Despite the wide variety of COMs, and the types of objects in which they are found, we still have not yet fully characterized the origins of COM emission observationally.

In low-mass star forming regions, COMs can be sublimated into the gas-phase from the ice mantles of dust grains with regions heated by the central protostars ($>100$ K). This zone is named the ``hot corino,'' a compact source ($\lesssim100$ au) at high temperature ($>$100 K) and density ($>10^7~{\rm cm^{-3}}$) \citep{ca12,ce22}.
In addition to the hot-corino origin, COMs are also found toward outflow cavity walls believed to be induced by shocks or protostellar UV irradiation \citep{jo04,ar08,su11,dr15,pa17}.
Furthermore, \citet{oy16} proposed that at the centrifugal radius where the infalling material from envelope lands on a protostellar disk, COMs can be liberated from dust mantles by weak accretion shocks in IRAS16293-2422A.
Similar scenarios are suggested with recent observations by
\citet{co18} and \citet{vas22} toward HH212 and BHB2007 11, respectively.
\citet{be20} categorized the COM emission in low-mass protostars into three groups (1) hot corinos, (2) outflows, and (3) accretion shock/disk atmosphere.

\begin{figure*}
\centering
\includegraphics[width=0.9\textwidth]{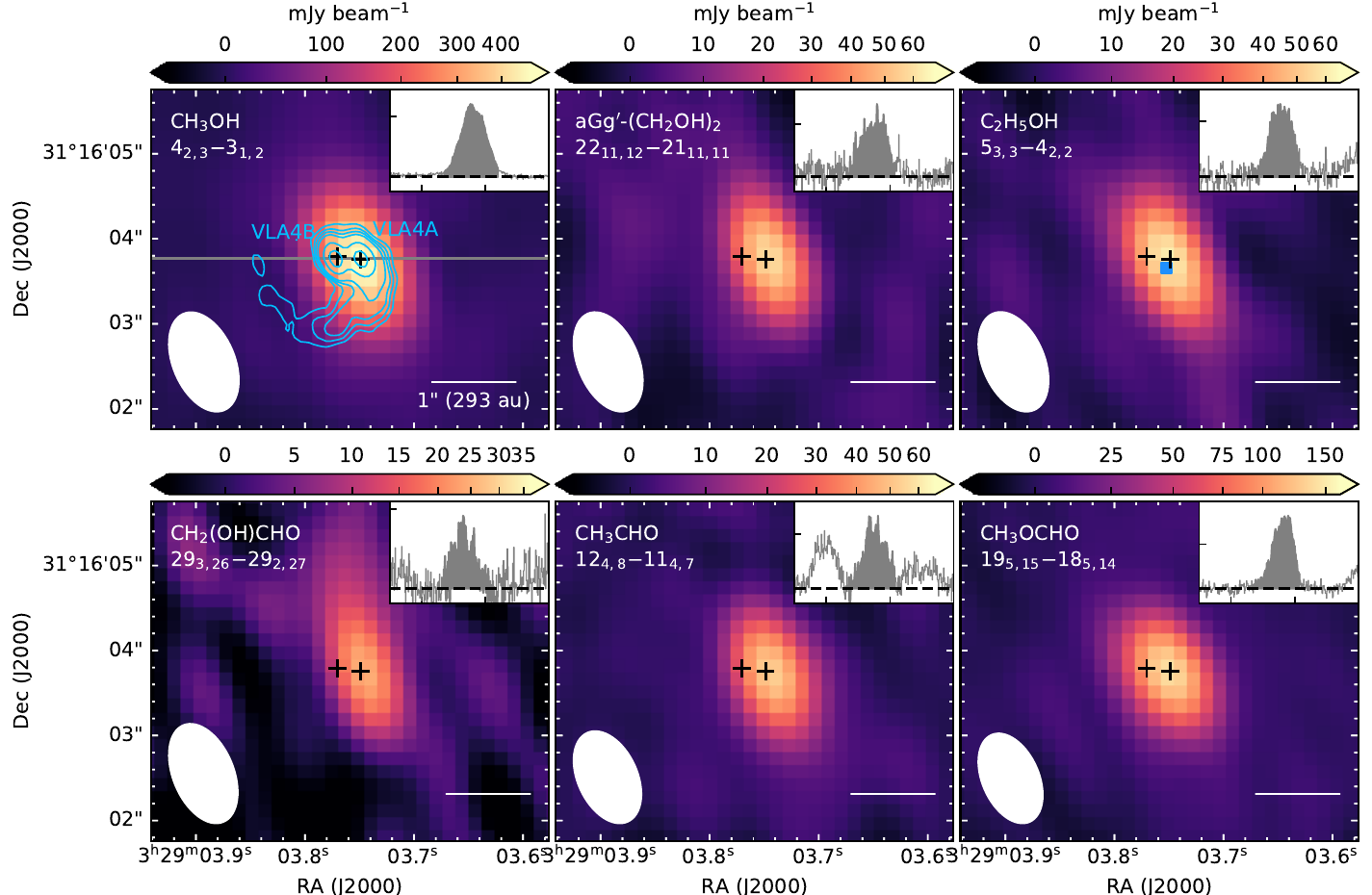}
\caption{Integrated intensity maps of a selected transition for each O-bearing COM. The upper right panel in each figure shows the spectra toward the center and the filled regions indicate the integration ranges for the zero-order moment maps, which is $3.5-10.7$ km s$^{-1}$, except for \ce{CH3OH} with $3.5-12.0$ km s$^{-1}$, for the zero-order moment maps. The blue contours in the top-left panel represent the ALMA 1.3 mm continuum emission \citep{to18} with levels of [3, 5, 7, 10, 30, 70]-$\sigma$. The horizontal grey line in the top left panel shows the PV cut used in Figure \ref{fig:pv}. The blue pixel at the top right panel indicate the pixel that is used to extract the spectra in this work.
}
\label{fig:mom0}
\end{figure*}

COMs are known to arise from SVS13A \citep{bi17,le18,bi19,be20,ya21,bi22a,bi22b,di22}, a Class I protobinary system located in the NGC 1333 cluster within the Perseus Molecular Cloud ($d=$293 pc, \citealt{or18}).
SVS13A contains two protostars, VLA4A and VLA4B, with a projected distance of $0\farcs3$ ($\sim$90 au) based on the continuum observations (first panel in Figure \ref{fig:mom0}, \citealt{an04,to16,to18,se18,ty20}).
With multiple-epoch observations, \citet{di22} derived a total mass of the binary to be 1.0$\pm$0.4 $M_\odot$ using orbital motions (see also \citealt{ma20}).
By modeling the kinematics of the line emission, they further estimate stellar masses of 0.27$\pm$0.10 $M_\odot$ for VLA4A and 0.60$\pm$0.20 $M_\odot$ for VLA4B.
The existence of a third hidden source at a distance of $\sim20-30$ au from VLA4B is implied by the wiggling jets seen at large scales \citep{le17}.
Through the dust continuum emission, at least two spirals are identified \citep{di22} and they were considered as part of a disk fragmenting due to gravitational instability \citep{to18}. However, recent observations found that the dust spiral is connected to larger scales (at least 700 au) through a streamer traced by DCN emission, which perhaps delivers material from the envelope to the central system \citep{hs23}.
Such infalling streamers at envelope scales are recently identified toward several protostellar systems, which are suggested to funnel material near or into smaller-scale disks \citep{pi20,al20,gi21,mu22,ca21,ga22,th22,va22,pi22}.
These infalling flows are believed to change the chemistry of the inner region (disk/envelope) by directly funneling material, induce shock gas, and/or indirectly trigger protostellar outbursts.
For SVS13A, the high luminosity $L_{\rm bol}=45.3~L_\odot$ suggests that it is undergoing a protostellar accretion outburst.
This makes SVS13A a good candidate to study the chemical inventory under the influence of accretion burst.

In this paper, we present new NOEMA data at 1.3 mm toward SVS13A. We select six oxygen-bearing (O-bearing) COMs (\ce{CH3OH}, aGg'-\ce{(CH2OH)2}, \ce{C2H5OH}, \ce{CH2(OH)CHO}, \ce{CH3CHO}, and \ce{CH3OCHO}) since they are found to be relatively bright with complex line profiles (multi-peaks) in our data; O-bearing COMs are also suggested to be more abundant relative to \ce{CH3OH} in shock regions \citep{cs18,cs19}. \ce{CH3OCH3} is also detected but in only a few blended transitions, preventing a detailed kinematic analysis (Appendix \ref{app:CH3OCH3}).
This work focuses on disentangling the kinematics of the protostellar system to unveil from which structures the COM emission originates, taking advantage of the high-spectral resolution and broadband capabilities of the IRAM-NOEMA PolyFiX correlator. 
In section \ref{sec:obs}, we describe the observations and calibration. The results are presented in \ref{sec:res}. In section \ref{sec:ana}, we detail the analysis and discussion in \ref{sec:dis}. The results are summarized in section \ref{sec:con}.

\section{Observations}
\label{sec:obs}
The observations were carried out with the NOrthern Extended Millimeter Array (NOEMA) of the Institut de Radioastronomie Millim\'{e}trique (IRAM).
It is part of the 
MPG-IRAM observing program PROtostars \& DIsks: Global Evolution (PRODIGE) (Project ID: L19MB002, PIs: P. Caselli and Th. Henning).
The observations are briefly described in \citet{hs23} that reports the data of \ce{C^18O} $J=2-1$, \ce{DCN} $J=3-2$, and \ce{CH3CN} $J=12_{\rm k}-11_{\rm k}$ (K=0--7).

In this paper, we present the data covering six O-bearing COMs, \ce{CH3OH}, aGg'-\ce{(CH2OH)2} (the most stable conformer of \ce{(CH2OH)2}, \citealt{ch01}), \ce{C2H5OH}, \ce{CH2(OH)CHO}, \ce{CH3CHO}, and \ce{CH3OCHO}. Here we describe the receiver setups with the PolyFiX correlator. The receiver contains four broadband dual polarization low-resolution windows and 39 narrowband high-resolution windows. The broadband windows cover
214.7$-$218.8 GHz, 218.8$-$222.8 GHz, 230.2$-$234.2 GHz, and 234.2$-$238.3 GHz with a channel width of 2 MHz ($\sim2.7$ km s$^{-1}$). The 39 narrowband windows are distributed within the mentioned ranges with a channel width of 62.5 kHz (Figure \ref{fig:spe}). The resulting spectral resolutions of the narrowband windows are $\sim0.078-0.086$ km s$^{-1}$.

Self-calibration was performed with solution intervals of 300 s, 135 s, and 45 s on the broadband data as in \citet{hs23}. The solutions are then applied to the broadband and narrowband data.
Imaging was done using the GILDAS/{\tt MAPPING} package\footnote{https://www.iram.fr/IRAMFR/GILDAS}.We use the {\tt clean} task with natural weighting in order to get the best sensitivity. This gives the beam sizes of $1\farcs11\times0\farcs67$ to $1\farcs24\times0\farcs73$ with a rms noise level of $\sim$0.5 K at a resolution of $\sim$0.8 km s${-1}$. Given the large number of channels in {both the broadband data and} the high-spectral resolution data sets, we do not clean using a support mask, and a shallow clean of the whole image with a 5$\sigma$ threshold is performed where $\sigma$ is the rms noise level. 
The single-point spectra toward the continuum peak (pixel at $\alpha$ = 3$^{\rm h}$29$^{\rm m}$ 03$^{\rm s}$.75, $\delta$ = 31$^{\rm d}$16$^{\rm m}$03$^{\rm s}$.73) are extracted for analysis in the paper.

\section{Results}
\label{sec:res}
\subsection{Line identification}
\label{sec:xclass}
\begin{figure*}
\centering
\includegraphics[width=1\textwidth]{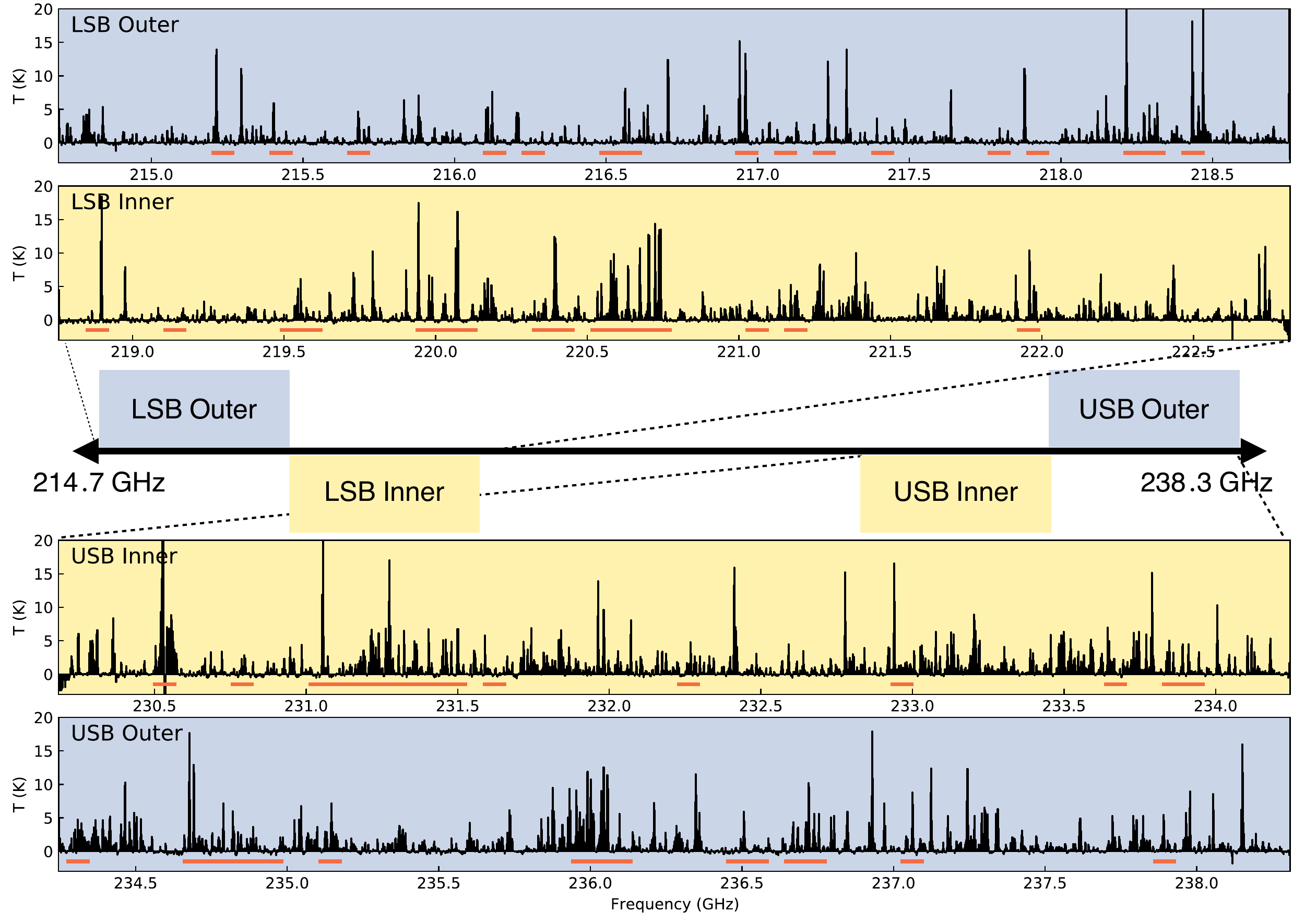}
\caption{Low-spectral-resolution spectrum toward the continuum peak of SVS13A from the PRODIGE NOEMA large program. The panels from the top to the bottom represent the spectra from the receiver bands as shown in the sketch in the middle. The horizontal orange bars represent locations of the high spectral-resolution windows.
}
\label{fig:spe}
\end{figure*}

Figure \ref{fig:spe} shows the full NOEMA spectrum from the broadband low-resolution data toward the continuum peak.
Line identification is first conducted using the lineID function of the eXtended CASA Line Analysis Software Suite (XCLASS) package, which automatically identifies lines, derives a quntitative description of each identified species, and return the corresponding physical parameters including the temperature, column density, source size, linewidth and systemic velocity (XCLASS, \citealt{mo17}) with the full broadband windows covering 16 GHz.
We will later disentangle the complicated line profiles using the high-spectral resolution data.

\subsection{Stacking of lines}
\label{sec:stack}
In order to increase the signal of the multiple peak structures, we stack the spectra from selected transitions for each COM; 
we manually select several transitions that are less contaminated by others. To be selected, the transition is expected to be broadly consistent with the XCLASS modeling result in amplitudes and share similar linewidths/profiles (section \ref{sec:xclass}). Column 7 in Table \ref{tab:tran_ch3oh} to \ref{tab:tran_ch3ocho} list the transitions which were stacked (see Appendix \ref{app:stack} and Figure \ref{fig:stack_all}).
The spectra are first interpolated using SciPy package \citep{scipy} to the same grid. Then, the spectra are normalized by the peak flux and averaged with a weight of inverse of square noise ($\sigma'$), i.e., the noise after normalization using NumPy \citep{numpy}.
\begin{equation}
    T_{\rm bri, stack} (v) = \frac{\sum_{i=0}^{N} T_{i} (v)\times \frac{1}{{\sigma'}_i^2}}{\sum_{i=0}^{N} \frac{1}{{\sigma'}_i^2}}.
\end{equation}
The stacked line profiles help to confirm multiple peaks in complex line profiles; we note that these lines have different opacities and energy levels so that their profiles are not necessary the same. 
For \ce{CH2(OH)CHO}, \ce{(CH2OH)2}, and \ce{C2H5OH}, there are limits of selected transitions or similar E$_{\rm up}$ ranged between 40-450 K for \ce{CH2(OH)CHO}, 110$-$266 K for aGg’-\ce{(CH2OH)2}, 23$-$409 K for \ce{C2H5OH}, and 129$-$207 K for CH3OCHO for \ce{CH3OCHO}. The stacked profile are broadly similar to the individual spectral profile (Figure \ref{fig:stack_all}). For \ce{CH3OH}, we split the transitions with E$_{\rm up}$ ranged 45$-$96 K and 775$-$802 K to stack. \ce{CH3CHO} have 13 selected transitions and are splitted into ones with 81$-$128 K and ones with 287$-$490 K.
Thus, as the stacked line profile reveal a complex line profile, to disentangle kinematic components from it requires a model.

\begin{figure*}
\centering
\includegraphics[width=1\textwidth]{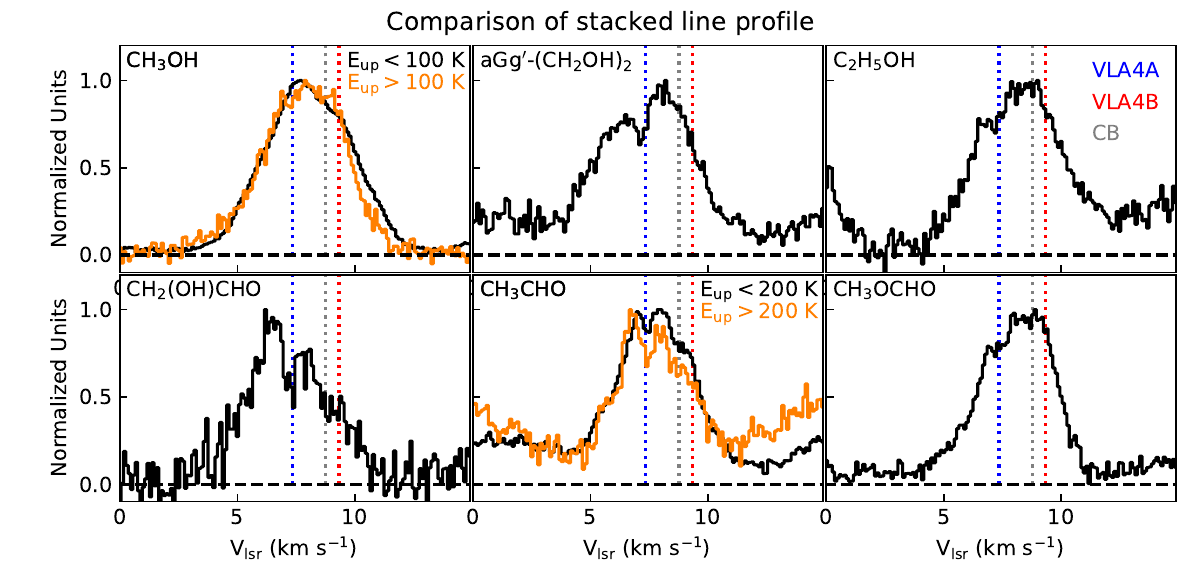}
\caption{Comparison of stacked line profiles of the six O-bearing COMs. The stacking is done using spectra normalized to their peak fluxes and weighting with the reciprocal of noise squared (Appendix \ref{app:stack}).
The vertical dashed lines indicate the central velocities of the disks of VLA4A (blue), 4B (red), and the circumbinary disk (grey) from \citet{di22}.}

\label{fig:stack}
\end{figure*}

Figure \ref{fig:stack} shows the stacked line profiles for the six selected O-bearing COMs toward the continuum peak.
It shows that different COMs have multiple peaks at different velocities, suggesting that they trace different kinematic components in the system.
In comparison to the systemic velocities of VLA4A/4B and the circumbinary disks, a dip is shown roughly at the velocity of VLA4A for the selected O-bearing COMs, with the exception of \ce{CH3OH}. For \ce{CH3CHO}, it seems a second dip at the systemic velocity of the circumbinary disk. These differences also support that these molecules trace different kinematic components.
These kinematic components are likely associated to the physical structures, for example, the disk, shocks, streamer, or the outflow.
aGg’-\ce{(CH2OH)2} shows a double peak structure likely tracing the rotating disk/envelope of VLA4A. 
This is consistent with the results from high-angular-resolution ALMA observations \citep{di22} that aGg’-\ce{(CH2OH)2} only traces the disk around VLA4A.
A double peak profile centered toward VLA4A is also seen in \ce{CH2(OH)CHO}. However, \ce{CH2(OH)CHO} is narrower and its blue-shifted peak is brighter while the opposite is true for aGg'-\ce{(CH2OH)2}.
On the other hand, \ce{C2H5OH}, \ce{CH3OCHO}, and \ce{CH3CHO} emission contains more than two peaks, which likely trace not only the VLA4A disk but also the circumbinary disk and/or VLA4B and perhaps streamers, shocks, inner envelopes, etc. These results are summarized in Table \ref{tab:comps}.

\begin{table*}
    \centering
    \caption{Emitting components based on spectra and location}
    \begin{tabular}{l cc cc cc cc cc}
    \hline\hline
&\multicolumn{2}{c}{aGg$'$-(CH$_2$OH)$_2$}
&\multicolumn{2}{c}{C$_2$H$_5$OH}
&\multicolumn{2}{c}{CH$_2$(OH)CHO}
&\multicolumn{2}{c}{CH$_3$CHO}
&\multicolumn{2}{c}{CH$_3$OCHO}
\\
\cmidrule(lr){2-3}
\cmidrule(lr){4-5}
\cmidrule(lr){6-7}
\cmidrule(lr){8-9}
\cmidrule(lr){10-11}
based on
& spectra
& spatial
& spectra
& spatial
& spectra
& spatial
& spectra
& spatial
& spectra
& spatial		\\
\hline
VLA4A       & \cmark & \cmark     & \cmark & \cmark     & \cmark  & \cmark     & \cmark  & \cmark     & \cmark & \cmark     \\
VLA4B       & \xmark & \xmark  & ? & \xmark   & \xmark  & \xmark   & ? & \xmark    & ? & \xmark \\
Circumbinary       & ? & \xmark  & \cmark & \cmark  & ?  & \xmark   & \cmark & \xmark & \cmark  & \cmark \\
\hline
    \end{tabular}\\
\tablefoot{The contributed components based on the stacked line profiles (``spectra'', section \ref{sec:stack}) and uv-domain Gaussian fitting (``spatial'', section \ref{sec:gau}).}
\label{tab:comps}
\end{table*}

\subsection{Gaussian widths and velocities}
To first examine the different physical structures a molecule traces, we conduct Gaussian fitting to the line profiles (narrowband) of each selected transition from the six O-bearing COMs. The transitions in use are those listed in \Cref{tab:tran_ch3oh,tab:tran_ch2oh2,tab:tran_c2h5oh,tab:tran_ch2ohcho,tab:tran_ch3cho,tab:tran_ch3ocho}.

\begin{figure}
\includegraphics[width=0.5\textwidth]{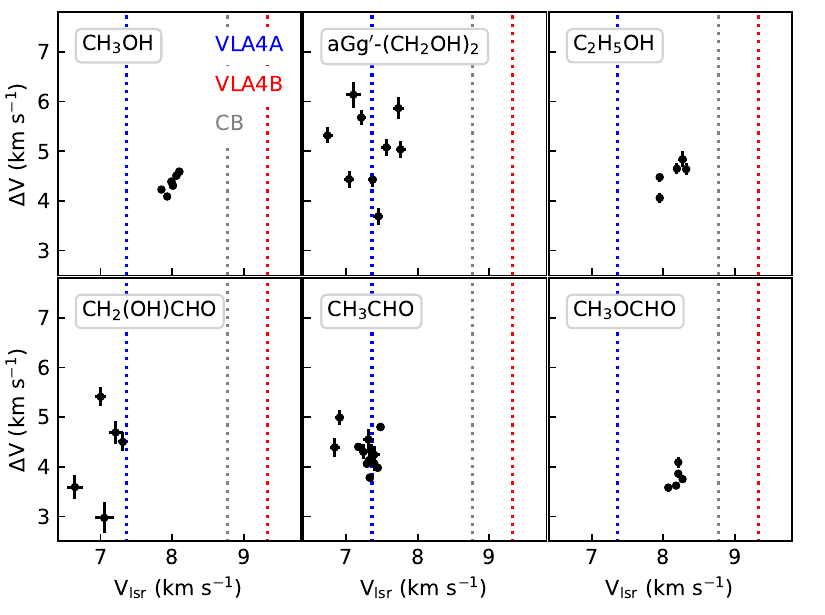}
\caption{Gaussian line width versus central velocity of the selected transitions from the six O-bearing COMs.  
The vertical dashed lines represent the central velocities of the disks of VLA4A (blue), 4B (red), and the circumbinary disk (grey).
The transitions in use are listed in \Crefrange{tab:tran_ch3oh}{tab:tran_ch3ocho}.
}
\label{fig:gau}
\end{figure}

Figure \ref{fig:gau} shows a comparison of the best-fit line widths and the central velocities of these spectral-line profiles with a single Gaussian. Although we have a limited number of non-contaminated transitions, it is clear that the linewidths and central velocities are different between the O-bearing COM species. This suggests that these COMs can be associated with different kinematic components in the protobinary system SVS13A. Considering the central velocities, it is likely that \ce{CH2(OH)CHO}, aGg'-\ce{(CH2OH)2}, and \ce{CH3CHO} emission is dominated by VLA4A while \ce{CH3OH}, \ce{C2H5OH}, and \ce{CH3OCHO} likely have significant emission coming from the circumbinary disk (and perhaps VLA4B).

\subsection{Spatial distribution of COM emission}
\label{sec:gau}
To derive properties of the gas traced by the COM emission, it is crucial to know the source size.
Figure \ref{fig:mom0} shows the integrated intensity maps of the six selected O-bearing COMs. All of them show point-source like structure given the spatial resolution of $\sim1\farcs2\times0\farcs7$ ($\sim300$ au).
The deconvolved source sizes from the uv-domain Gaussian fitting ($\lesssim0\farcs3$, 90 au) are smaller than the beam size, suggesting that they are point sources at the current resolution. This gives an upper limit of the emitting regions of $\sim90$ au. More discussions on the source sizes are provided in sections \ref{sec:oneGau} and \ref{sec:multiGau}. The compact sources might rule out the possibility that the emission comes from larger-scale structures, such as irradiated cavity walls carved by the bipolar outflows.

\begin{figure*}
\centering
\includegraphics[width=0.8\textwidth]{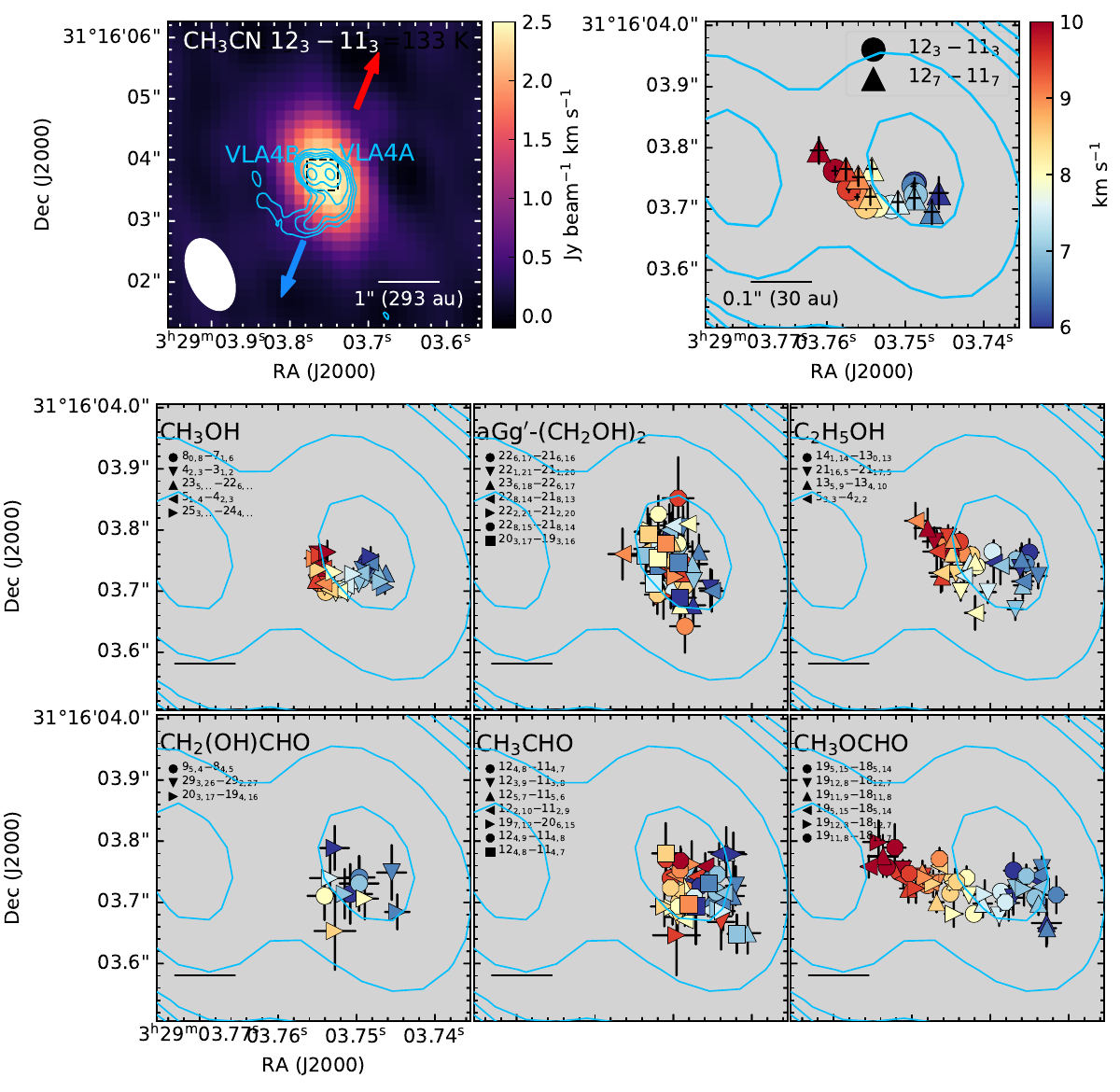}
\caption{Central positions of selected COM transitions at different velocities. The top two panels show the integrated intensity map (left) and central positions (right) from \ce{CH3CN} \citep{hs23} as a reference for the O-bearing COMs.
The contours show the 1.3 mm continuum emission from \citet{to18}.
Different line transitions are marked in different symbols (\Cref{tab:tran_ch3oh,tab:tran_ch2oh2,tab:tran_c2h5oh,tab:tran_ch2ohcho,tab:tran_ch3cho,tab:tran_ch3ocho}).
The scale bar is $0\farcs1$ for the bottom six panels. The beam sizes for these COM emissions are depending on the line frequency but is the same to as \ce{CH3CN} ($1\farcs2\times0\farcs7$) within 10\%.}
\label{fig:uv_cen}
\end{figure*}

The peak positions of line emission are used to indicate the most likely location of the emitting gas at specific velocity, especially for optically thin lines \citep{sa87,ha13}. Toward SVS13A, \citet{hs23} conducted a uv-domain Gaussian fitting to \ce{CH3CN} emission using a channel width of 0.5 km s$^{-1}$, and found a velocity gradient from west to east (Figure \ref{fig:uv_cen}, top panel); the visibility data is first resampled to a width of 0.5 km s$^{-1}$, and 
{\tt uv\_fit} task in {\tt MAPPING} is used to conduct the 2D Gaussian fit for each channel.
We applied the same technique to six O-bearing COMs in SVS13A; this was done with the selected transitions (From \Crefrange{tab:tran_ch3oh}{tab:tran_ch3ocho}). As a result, we reveal the spatial weighting centers of each velocity for each transition.

The O-bearing lines have lower S/N ratios compared to the \ce{CH3CN} emission, especially in the red or blue shifted wings.
For uv-plane fitting of the O-bearing COM emission, we only consider those data points with S/N$>$7 and position uncertainty $<0\farcs5$.
Figure \ref{fig:uv_cen} shows the fitted central positions.
It is noteworthy that, for each molecule, although the transitions have different upper energy levels and opacities (\Cref{tab:tran_ch3oh,tab:tran_ch2oh2,tab:tran_c2h5oh,tab:tran_ch2ohcho,tab:tran_ch3cho,tab:tran_ch3ocho}), the spatial distributions are similar.
It is clear that the emission of these O-bearing COMs originates from different locations in the system. \ce{CH3OH}, \ce{C2H5OH}, and \ce{CH3OCHO} show a velocity gradient with the direction consistent with that of \ce{CH3CN}. 
These three molecules also share similar systemic velocity $\sim8.1$ km s$^{-1}$ from the line profile fitting (Figure \ref{fig:gau}).
However, the red-shifted end of \ce{CH3OCHO} extends close to VLA4B while the \ce{CH3OH} is relatively compact and likely dominated by VLA4A.
For aGg'-\ce{(CH2OH)2}, \ce{CH2(OH)CHO}, and \ce{CH3CHO}, the line emission most likely comes from VLA4A. This in general agrees with the results from line stacking, except for \ce{CH3CHO} which has $3-4$ components (Figure \ref{fig:stack}). However, considering their Gaussian peak velocity, these molecules have emission dominated by VLA4A (Figure \ref{fig:gau}). We summarize these in Table \ref{tab:comps}.

\section{Analysis}
\label{sec:ana}
\subsection{One-component LTE Gaussian model}
\label{sec:oneGau}
We apply radiative transfer modeling to the emission of selected transitions of the six O-bearing COMs.
Though multiple peaks are found in the line profiles from the high-spectral resolution data (Figure \ref{sec:stack}),
we first perform LTE (Local Thermodynamic Equilibrium) modeling of the spectra with one Gaussian velocity component.
We use the LTE approach as the collisional rate coefficients of COMs are only available in a few COMs (e.g., Leiden Atomic and Molecular Database
or LAMDA\footnote{\url{https://home.strw.leidenuniv.nl/~moldata/}}). 
This one component model provides the first-order estimate to the physical conditions of the gas traced by each COM molecule; we will use multiple Gaussian component modeling to decompose the complex structures in Section \ref{sec:multiGau}.
Non-LTE analysis has been done for \ce{CH3OH} (\citealt{bi17}: $n_{\rm H_2}>10^8$ cm$^{-3}$) and \ce{CH3CN} (\citealt{hs23}: $n_{\rm H_2}\sim6.2\times10^6$ cm$^{-3}$ and $n_{\rm H_2}>1.2\times10^8$ cm$^{-3}$ with two components). 
These analyses, for the only two COMs with known collision coefficients, suggest that COMs come from the central dense regions, implying that LTE analysis is reasonable for other COMs.

We applied the model to fit the high-spectral-resolution spectra extracted from the position of the peak continuum emission.
The NOEMA observations at 1.3 mm cover $\sim$200-1400 transitions, depending on the species (Figure \ref{fig:spe}). At the same time, line contamination from nearby transitions becomes a severe issue to model the line profiles (Figure \ref{fig:spe}).
We, therefore, check the line profiles from all transitions covered in the high-spectral resolution data for each molecule.
We select as many transitions as possible with a range of line strengths and upper energy levels to better constrain the physical conditions (Tables \ref{tab:tran_ch3oh} to \ref{tab:tran_ch3ocho}). 
The selection process is done by iteratively checking the model and line profiles of all transitions located in the narrow band windows. 
Based on the XCLASS model (section \ref{sec:xclass}, fitting the low-spectral-resolution data with the latest and most complete list of molecules), a transition which is contaminated by different molecules is first removed.
A line emission much higher than the model profile may be also considered as contaminated by unidentified lines.
In the other word, the final (one-velocity-component) model of the high-spectral resolution data is broadly consistent with the XCLASS model considering more channels but fewer transitions.
Line contamination can also be treated by masking the specific, overlapping velocity ranges so that more transitions can be included (see Appendix \ref{app:fitting}: \Cref{fig:Fitting_ch3oh,fig:Fitting_ch2oh2,fig:Fitting_c2h5oh,fig:Fitting_ch2ohcho,fig:Fitting_ch3cho,fig:Fitting_ch3ocho}).
Unlike line stacking (section \ref{sec:stack}), the fit includes line profiles with several overlapping transitions from the same molecule (see below).
In addition, undetected transitions can be included as they help to constrain the physical conditions, especially when the transition has a high Einstein coefficient and/or low upper-energy level.

We construct the LTE-radiative transfer model to fit the line profiles of the selected transitions. The model includes five free parameters which are column density ($N_{\rm tot}$), excitation temperature ($T_{\rm ex}$), central velocity ($V_{\rm LSR}$), linewidth ($\Delta V$), and source size ($\Omega_S$).
We note here the linewidth is set to a free parameter as one value for all transitions. This intrinsic linewidth is for $\tau(v)$ so that it should be smaller than the real linewidth in the optically thick transitions.
The line opacity as a function of velocity (frequency) is first constructed following the XCLASS manual\footnote{http://cassis.irap.omp.eu/docs/RadiativeTransfer.pdf} \citep{mo17}:
\begin{equation}
    \tau (v)=\sum_{\rm i} \tau_{\rm i,peak} \exp(-\frac{(v-V_{\rm LSR})^2}{2\sigma^2}),
\end{equation}
where
\begin{equation}
    \tau_{\rm peak}=\frac{A_{\rm ul} c^3}{8\pi \nu^3 \Delta V \frac{(\pi ln2)^{1/2}}{2}} N_{\rm u} (e^{h\nu/kT_{\rm ex}}-1),
    \label{eq:tau}
\end{equation}
for which $i$ indicates the transition in use, $\sigma\sim\frac{\Delta V}{2\sqrt{2\ln{2}}}$, $A_{\rm ul}$ is the Einstein coefficient, c is light speed, and $\nu$ is the rest frequency of the transition from CDMS\footnote{https://cdms.astro.uni-koeln.de/cdms/portal/} \citep{mu05,en16}.
$N_{\rm u}$ is the column density at the upper energy state and can be expressed as,
\begin{equation}
    N_{\rm u}=N_{\rm tot}\frac{g_{\rm u}}{Q(T_{\rm ex})} e^{-E_{\rm u}/kT_{\rm ex}},
\end{equation}
where $g_{\rm u}$ is the degeneracy, $Q(T_{\rm ex})$ is the partition function, and $E_{\rm u}$ is the upper energy level.
Finally, the line intensity is constructed as
\begin{equation}
    I_{\nu} (v)=\frac{\Omega_S}{\Omega_B} (J_{\nu}(T_{\rm ex})-J_\nu(T_{\rm bg})) (1-\exp(-\tau (v))),
\label{eq:rad}
\end{equation}
where $\Omega_S$ and $\Omega_B$ are the source size and beam size, respectively, and $T_{\rm bg}=2.7 $K.
We note that $\Omega_B=1\farcs2\times0\farcs7$ as we first scale the intensity of the observed spectra into this beam size for convenience.

We conduct the fitting using the Scipy {\tt curve\_fit} method\footnote{https://scipy.org/citing-scipy/} to estimate initial parameter values.
Then, {\tt emcee} \citep{fo13} is used to perform MCMC sampling using the parameters from {\tt curve\_fit} estimates with uniform distribution for priors.
The resulting best-fit parameters are listed in Table \ref{tab:mcmc} denoted with the component as single.
The fitted parameters and their uncertainties are taken with relevant 16\%, 50\%, and 84\% quantiles \citep{ho18}.
We find that each O-bearing molecule probes gas with different physical conditions.
The excitation temperatures are from $\sim145$ K (\ce{C2H5OH}) to $\sim240$ K (aGg'-\ce{(CH2OH)2}) and the source sizes vary from $\sim0\farcs10$ (aGg'-\ce{(CH2OH)2}) to $\sim0\farcs28$ (\ce{CH2(OH)CHO}). This again suggests that these COMs come from different components/regions of this protobinary system.

\Cref{fig:Fitting_ch3oh,fig:Fitting_ch2oh2,fig:Fitting_c2h5oh,fig:Fitting_ch2ohcho,fig:Fitting_ch3cho,fig:Fitting_ch3ocho} show the best-fit in use (\Cref{tab:tran_ch3oh,tab:tran_ch2oh2,tab:tran_c2h5oh,tab:tran_ch2ohcho,tab:tran_ch3cho,tab:tran_ch3ocho}).
Transitions with optically thick emission are crucial to break the degeneracy between source size and column density (e.g., $\Omega_S$ and $\tau$ in equation \ref{eq:rad}).
On the other hand, to determine the temperature, transitions covering a broad range of energy levels are necessary.
Figure \ref{fig:trans} shows the $\tau$ of the best-fit model as a function of E$_{\rm u}$ for the transitions in use. We use this figure to evaluate the the fitting results. For example, \ce{CH2(OH)CHO} fitting is done using only optically thin lines ($\tau\lesssim$0.13);
in such a case, equation \ref{eq:rad} is approximately $I_{\nu} (v)\propto{\Omega_S}\tau (v)$. 
Hence, $\tau$, which is proportional to the column density of the energy state (\Cref{eq:tau}), is degenerate with source size; this results in an unconstrained posterior distribution in the MCMC sampling.
We note that although the column density cannot be constrained, the intensity ratios can still reflect the temperature in the optically thin case. In addition, the total amount of molecules can be estimated (i.e., column 8 in Table \ref{tab:mcmc}).

\subsection{Multi-velocity component LTE Gaussian model}
\label{sec:multiGau}
The line profiles of these O-bearing COMs usually contain multiple peaks (Figure \ref{fig:stack}). This suggests that one COM can trace or several kinematic components, and different COMs can trace different regions in SVS13A. The multiple components do not trace the same material. Therefore, we perform multiple velocity components fitting to disentangle the physical conditions traced by the line profiles. For \ce{CH3OH}, we do not conduct multiple component fitting as only a few transitions are observed and the energy levels of E$_{\rm up}$ from $100-700$ K are not present in our setup.

The multiple-velocity-component model setup is the same as that of the one-component model (section \ref{sec:oneGau}) but with more Gaussian components.
The fitting is using the same spectra with selected transitions and mask to the one-velocity-component case (section \ref{sec:oneGau}).
Each component has its own physical parameters so that one adds five additional free parameters (
$T_{\rm ex}$, $\log N_{\rm tot}$, $\Delta V$, $V_{\rm LSR}$, size) for each additional velocity component added to the model.
We assume these components are spatially separated and do not interact.

To decide how many velocity components are needed to fit the line emission, we employ the Akaike Information Criterion (AIC) $AIC=2k+\chi^2+C$ where $k$ is the number of the free parameters and $\chi$ is the classical chi-squared statistic (see also \citealt{ch20,va22}). 
In our case, $k$ is equal to 5 times the number of components.
Thus, by adding one velocity component, $\chi^2$ needs to decrease by 10 to compensate AIC. A small AIC is in favor for the model.
We conduct $\chi^2$ fitting starting with one component, and adding component one by one until the $\Delta AIC$ does not significantly change (Table \ref{tab:aic}).
The details are described in Appendix \ref{app:fitting}.
This progress to decide the number of velocity components is done using gradient based optimization (Scipy {\tt curve\_fit}) with less computational expense than MCMC method.

\begin{figure*}
\centering
\includegraphics[width=1\textwidth]{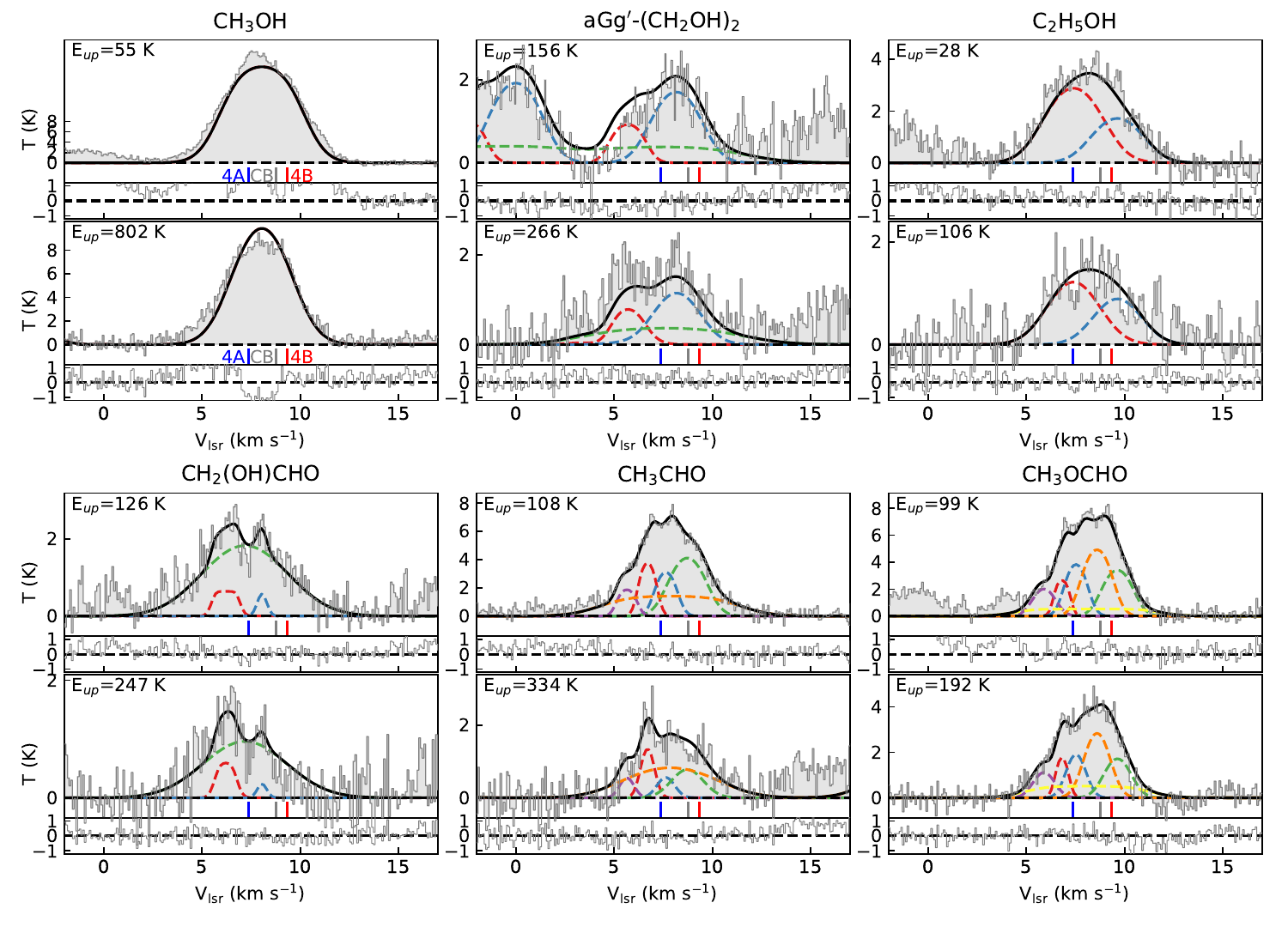}
\caption{Spectral profiles of two selected transitions for each O-bearing COM. The black curve shows the best-fit model with multiple kinematic components and the colored dashed lines represent the contribution from each component. The blue, red, and grey bars in the bottom indicate the systemic velocities of VLA4A, VLA4B, and the circumbinary disk, respectively. The insets at the bottom of each panel show the residual from the best-fit model.
}
\label{fig:fitting}
\end{figure*}

After deciding the number of components, we then perform the MCMC sampling. 
The Markov-chains start from positions surrounding the best-fitting model in high-dimensional space with uniform priors; tens of walkers (depending on the $k$) were running simultaneously using the ``Stretch Moves'' in {\tt emcee}\citep{go10,fo13}. 
The initial values are set to the nearby the best-fit model from {\tt curve\_fit}.
The same as the one-component model (section \ref{sec:oneGau}), the best-fit parameters and the uncertainties are taken with 16\%, 50\%, and 84\% quantiles \citep{ho18}. Table \ref{tab:mcmc} shows these results and Figure \ref{fig:fitting} shows the best-fit models for each COM in two selected transitions which have different upper energy levels and relatively high S/N among the transitions in use for fitting.
We find that for a given species, the line emission can consist of several components with different temperatures.
The temperatures are different from the single-component fitting by $\lesssim30\%$ except for aGg'-\ce{(CH2OH)2}, for which the three-component fitting yields temperatures of 180$-$470 K in comparison to the 240 K by the one-component model.

It is important to discuss how opacity affects the fitting process.
In the fitting process, the line interaction between components is not included, i.e., each velocity component is assumed to be spatially separated.
In \ce{C2H5OH} and \ce{CH3OCHO}, the velocity components are relatively offset from each other so that the opacity effects are small in the overlapping region.
In aGg'-\ce{(CH2OH)2}, \ce{CH3CHO}, and \ce{CH2(OH)CHO}, a broad-line component is found. This component overlaps with the others. For \ce{CH2(OH)CHO}, based on the single component fitting, the line emission of the broad-line component is most likely optically thin in the best-fit model (Figure \ref{fig:trans}).
For aGg'-\ce{(CH2OH)2} and \ce{CH3CHO}, it is likely that their broad-line components include some optically-thick transitions in the fitting. These broad-line components with higher column density and small source sizes might be associated with the central compact regions (see also section \ref{sec:pv}). If this is the case, this compact emission is behind the extended sources. Thus, the physical conditions of the spatially-compact broad-line components are inaccurate while the remaining components are less affected.
This is however the limit of the data set. Higher angular resolution data is required to resolve these issues.

\begin{figure*}
\centering
\includegraphics[width=1\textwidth]{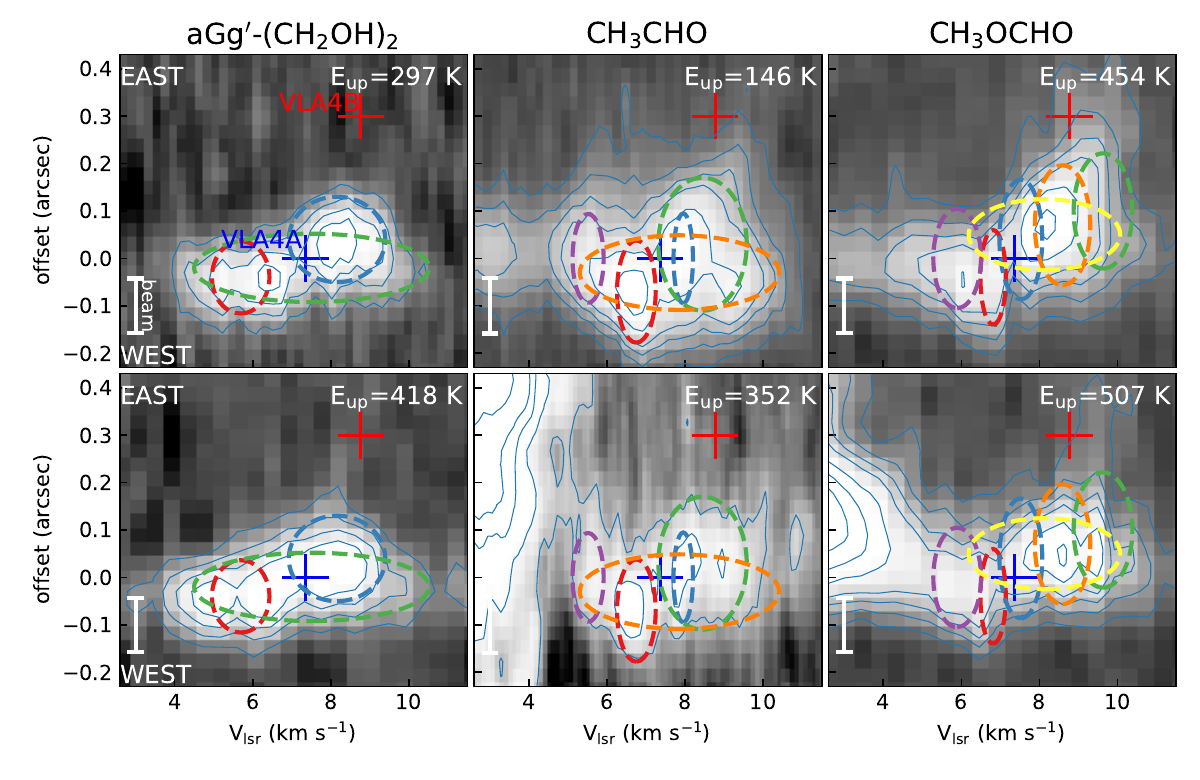}
\caption{PV diagrams of aGg'-\ce{(CH2OH)2}, \ce{CH3CHO}, and \ce{CH3OCHO} from high-angular-resolution ALMA data \citep{di22}. The color dashed ellipses represent each kinematic component from the best-fit model (with corresponding color as Figure \ref{fig:fitting}). The blue and red crosses indicate the positions and velocities of VLA4A and VLA4B, respectively. The white bar in the lower left corner shows the beam size along the PV cut. 
NOEMA fits on top of the higher-resolution ALMA data match the structure, validating the NOEMA fit results.
}
\label{fig:pv}
\end{figure*}

\begin{table*}
\setlength{\tabcolsep}{3.5pt}
    \centering
    \def\arraystretch{1.2}
    \caption{Physical conditions of COMs from MCMC sampling}
    \begin{tabular}{l ccccccccc}
    \hline\hline
&
& $T_{\rm ex}$
& $\log N_{\rm tot}$
& $\Delta V$
& $V_{\rm LSR}$
& D (diameter)
& $N_{\rm tot}\times {\rm area}^{a}$
& $\frac{N_{\rm X}}{N_{\rm \ce{CH3OH}}}^{b}$
& $\frac{[\rm X]}{[\rm \ce{CH3OH}]}^{c}$\\
COM
& component
& (K)
& (cm$^{-2}$)
& (km s$^{-1}$)
& (km s$^{-1}$)
& (arcsec)
& (cm$^{-2}$ arcsec$^2$)
&
&\\
\hline
CH$_3$OH        & single        & 229.3$^{+0.72}_{-0.64}$       & 18.9$^{+0.01}_{-0.01}$        & 3.12$^{+0.01}_{-0.01}$      & 8.05$^{+0.01}_{-0.01}$ & 0.279$^{+0.001}_{-0.001}$     & 1.95$^{+0.01}_{-0.01}\times 10^{18}$  & 1     & 1\\
\hline
aGg$'$-(CH$_2$OH)$_2$   & single        & 240.6$^{+14.0}_{-13.15}$      & 18.17$^{+0.04}_{-0.04}$       & 3.85$^{+0.06}_{-0.06}$       & 7.55$^{+0.02}_{-0.02}$        & 0.1$^{+0.003}_{-0.003}$       & 4.69$^{+0.53}_{-0.53}\times 10^{16}$  & 0.186 & 0.024\\
\hline
aGg$'$-(CH$_2$OH)$_2$   & comp 1        & 278.7$^{+62.06}_{-46.59}$     & 18.05$^{+0.18}_{-0.16}$       & 1.48$^{+0.07}_{-0.07}$       & 5.68$^{+0.08}_{-0.08}$        & 0.057$^{+0.006}_{-0.006}$     & 1.16$^{+0.54}_{-0.54}\times 10^{16}$  & -     & 0.036\\
        &comp 2 & 184.9$^{+16.46}_{-15.28}$     & 17.68$^{+0.07}_{-0.08}$       & 2.5$^{+0.12}_{-0.13}$ & 8.16$^{+0.05}_{-0.05}$       & 0.112$^{+0.005}_{-0.005}$     & 1.89$^{+0.38}_{-0.38}\times 10^{16}$  & -     & -\\
        &comp 3 & 471.4$^{+201.4}_{-118.9}$     & 19.22$^{+0.36}_{-0.28}$       & 6.03$^{+0.46}_{-0.42}$        & 7.52$^{+0.2}_{-0.17}$        & 0.027$^{+0.006}_{-0.006}$     & 3.92$^{+3.72}_{-3.72}\times 10^{16}$  & -     & -\\
\hline
C$_2$H$_5$OH    & single        & 97.53$^{+1.81}_{-1.74}$       & 17.44$^{+0.02}_{-0.02}$       & 3.88$^{+0.04}_{-0.04}$      & 8.24$^{+0.02}_{-0.02}$ & 0.254$^{+0.005}_{-0.004}$     & 5.56$^{+0.37}_{-0.37}\times 10^{16}$  & 0.035 & 0.029\\
\hline
C$_2$H$_5$OH    & comp 1        & 87.03$^{+2.75}_{-2.76}$       & 17.3$^{+0.04}_{-0.04}$        & 3.12$^{+0.12}_{-0.11}$      & 7.53$^{+0.1}_{-0.1}$   & 0.233$^{+0.008}_{-0.008}$     & 3.4$^{+0.37}_{-0.37}\times 10^{16}$   & -     & 0.032\\
        &comp 2 & 142.7$^{+24.51}_{-15.36}$     & 17.7$^{+0.19}_{-0.14}$        & 2.75$^{+0.16}_{-0.13}$        & 9.72$^{+0.1}_{-0.1}$ & 0.134$^{+0.019}_{-0.021}$     & 2.84$^{+1.52}_{-1.52}\times 10^{16}$  & -     & -\\
\hline
CH$_2$(OH)CHO   & single        & 151.9$^{+2.4}_{-2.97}$        & 16.86$^{+0.29}_{-0.36}$       & 4.49$^{+0.12}_{-0.12}$      & 7.04$^{+0.04}_{-0.04}$ & 0.268$^{+0.132}_{-0.072}$     & 1.64$^{+2.11}_{-2.11}\times 10^{16}$  & 0.009 & 0.008\\
\hline
CH$_2$(OH)CHO   & comp 1        & 115.3$^{+31.2}_{-20.18}$      & 17.43$^{+0.22}_{-0.27}$       & 0.81$^{+0.18}_{-0.13}$      & 6.22$^{+0.07}_{-0.06}$ & 0.07$^{+0.011}_{-0.009}$      & 4.09$^{+2.84}_{-2.84}\times 10^{15}$  & -     & 0.010\\
        &comp 2 & 134.8$^{+109.8}_{-44.95}$     & 15.58$^{+0.75}_{-0.58}$       & 0.68$^{+1.93}_{-0.27}$        & 8.09$^{+0.36}_{-0.11}$       & 0.155$^{+0.097}_{-0.08}$      & 2.88$^{+6.16}_{-6.16}\times 10^{14}$  & -     & -\\
        &comp 3 & 149.7$^{+3.74}_{-3.79}$       & 16.98$^{+0.24}_{-0.19}$       & 4.78$^{+0.18}_{-0.17}$        & 7.18$^{+0.07}_{-0.07}$       & 0.223$^{+0.051}_{-0.05}$      & 1.49$^{+1.07}_{-1.07}\times 10^{16}$  & -     & - \\
\hline
CH$_3$CHO       & single        & 154.6$^{+1.33}_{-1.25}$       & 17.12$^{+0.01}_{-0.01}$       & 3.41$^{+0.01}_{-0.02}$      & 7.76$^{+0.01}_{-0.01}$ & 0.252$^{+0.002}_{-0.002}$     & 2.62$^{+0.06}_{-0.06}\times 10^{16}$  & 0.017 & 0.013\\
\hline
CH$_3$CHO       & comp 1        & 143.3$^{+5.38}_{-4.99}$       & 16.83$^{+0.04}_{-0.04}$       & 0.99$^{+0.04}_{-0.04}$      & 6.74$^{+0.02}_{-0.02}$ & 0.16$^{+0.007}_{-0.008}$      & 5.42$^{+0.74}_{-0.74}\times 10^{15}$  & -     & 0.019\\
        &comp 2 & 105.4$^{+12.32}_{-10.74}$     & 16.22$^{+0.08}_{-0.1}$        & 0.5$^{+0.03}_{-0.03}$ & 7.95$^{+0.01}_{-0.01}$       & 0.127$^{+0.008}_{-0.007}$     & 8.4$^{+2.13}_{-2.13}\times 10^{14}$   & -     & -\\
        &comp 3 & 116.5$^{+2.58}_{-2.63}$       & 16.73$^{+0.03}_{-0.03}$       & 2.29$^{+0.06}_{-0.06}$        & 8.43$^{+0.04}_{-0.04}$       & 0.242$^{+0.004}_{-0.004}$     & 9.91$^{+0.84}_{-0.84}\times 10^{15}$  & -     & -\\
        &comp 4 & 133.9$^{+12.2}_{-10.87}$      & 16.58$^{+0.05}_{-0.05}$       & 0.8$^{+0.04}_{-0.04}$ & 5.51$^{+0.03}_{-0.03}$       & 0.123$^{+0.005}_{-0.005}$     & 1.8$^{+0.27}_{-0.27}\times 10^{15}$   & -     & -\\
        &comp 5 & 182.0$^{+11.99}_{-10.85}$     & 18.09$^{+0.07}_{-0.07}$       & 5.14$^{+0.17}_{-0.16}$        & 7.85$^{+0.06}_{-0.06}$       & 0.069$^{+0.004}_{-0.004}$     & 1.82$^{+0.36}_{-0.36}\times 10^{16}$  & -     & -\\
\hline
CH$_3$OCHO      & single        & 173.6$^{+2.05}_{-2.03}$       & 17.44$^{+0.01}_{-0.01}$       & 3.53$^{+0.01}_{-0.01}$      & 8.17$^{+0.01}_{-0.01}$ & 0.348$^{+0.002}_{-0.002}$     & 1.05$^{+0.03}_{-0.03}\times 10^{17}$  & 0.035 & 0.054\\
\hline
CH$_3$OCHO      & comp 1        & 134.4$^{+14.87}_{-12.21}$     & 17.24$^{+0.1}_{-0.09}$        & 0.65$^{+0.04}_{-0.04}$      & 6.81$^{+0.02}_{-0.02}$ & 0.136$^{+0.015}_{-0.015}$     & 1.01$^{+0.33}_{-0.33}\times 10^{16}$  & -     & 0.079\\
        &comp 2 & 111.3$^{+6.48}_{-6.1}$        & 17.01$^{+0.05}_{-0.06}$       & 1.13$^{+0.07}_{-0.07}$        & 7.52$^{+0.04}_{-0.05}$       & 0.214$^{+0.012}_{-0.012}$     & 1.47$^{+0.25}_{-0.25}\times 10^{16}$  & -     & -\\
        &comp 3 & 116.2$^{+7.87}_{-7.72}$       & 17.15$^{+0.05}_{-0.06}$       & 1.54$^{+0.08}_{-0.07}$        & 9.62$^{+0.07}_{-0.07}$       & 0.202$^{+0.014}_{-0.013}$     & 1.8$^{+0.34}_{-0.34}\times 10^{16}$   & -     & -\\
        &comp 4 & 168.6$^{+18.13}_{-15.37}$     & 17.11$^{+0.13}_{-0.16}$       & 1.25$^{+0.1}_{-0.09}$ & 5.88$^{+0.06}_{-0.05}$       & 0.16$^{+0.021}_{-0.014}$      & 1.04$^{+0.46}_{-0.46}\times 10^{16}$  & -     & -\\
        &comp 5 & 133.8$^{+7.49}_{-6.81}$       & 17.36$^{+0.05}_{-0.04}$       & 1.45$^{+0.06}_{-0.06}$        & 8.59$^{+0.05}_{-0.05}$       & 0.209$^{+0.012}_{-0.014}$     & 3.13$^{+0.54}_{-0.54}\times 10^{16}$  & -     & -\\
        &comp 6 & 179.3$^{+19.18}_{-16.13}$     & 18.93$^{+0.09}_{-0.08}$       & 3.81$^{+0.19}_{-0.19}$        & 8.15$^{+0.11}_{-0.11}$       & 0.051$^{+0.005}_{-0.004}$     & 7.03$^{+1.88}_{-1.88}\times 10^{16}$  & -     & -\\
\hline
    \end{tabular}
\tablefoot{For each molecule, we provide the best-fit parameters for the one-component LTE model (section \ref{sec:oneGau}) denoted as single in the second column and for the multi-velocity component LTE model (section \ref{sec:multiGau}) denoted as comp x in the second column.\\
    \tablefoottext{a}{The area is derived as $\pi r^2$ where $r$ is $D/2$.}\\
    \tablefoottext{b}{The column density ratio with respect to \ce{CH3OH}.}\\
    \tablefoottext{c}{The abundance ratio with respect to \ce{CH3OH} derived from the $N_{\rm tot}\times {\rm area}$, i.e., column 8. For multiple-velocity case, the $N_{\rm tot}\times {\rm area}$ from each component are sum up.}
}

\label{tab:mcmc}
\end{table*}

\begin{table*}
    \centering
    \caption{Multi-component model comparison}
    \begin{tabular}{l cc cc cc cc cc}
    \hline\hline
&\multicolumn{2}{c}{aGg$'$-(CH$_2$OH)$_2$}
&\multicolumn{2}{c}{C$_2$H$_5$OH}
&\multicolumn{2}{c}{CH$_2$(OH)CHO}
&\multicolumn{2}{c}{CH$_3$CHO}
&\multicolumn{2}{c}{CH$_3$OCHO}
\\
\cmidrule(lr){2-3}
\cmidrule(lr){4-5}
\cmidrule(lr){6-7}
\cmidrule(lr){8-9}
\cmidrule(lr){10-11}
Num. of components
& $\chi^2$
& $\Delta AIC$	
& $\chi^2$
& $\Delta AIC$	
& $\chi^2$
& $\Delta AIC$	
& $\chi^2$
& $\Delta AIC$	
& $\chi^2$
& $\Delta AIC$	\\
\hline
1       & 12993 & 0     & 39843 & 0     & 4911  & 0     & 11379 & 0     & 11323 & 0     \\
2       & 12865 & -118  & 39774 & -59   & 4888  & -13   & 10840 & -529  & 10428 & -885  \\
3       & 12773 & -82   & (39751)       & (-13) & 4824  & -54   & 10443 & -387  & 10234 & -184  \\
4       & (12753)       & (-10) & -     & -     & (4802)        & (-12) & 10329 & -104  & 9991  & -233  \\
5       & -     & -     & -     & -     & -     & -     & 10083 & -236  & 9713  & -268  \\
6       & -     & -     & -     & -     & -     & -     & (10054)       & (-19) & 9563  & -140  \\
7       & -     & -     & -     & -     & -     & -     & -     & -     & (9534)        & (-19) \\
\hline
    \end{tabular}\\
\tablefoot{$\Delta$AIC is in comparison with the previous level. The brackets around the final rows indicate the stop point at which $\Delta$AIC (as well as $\chi^2$) do not significantly change; therefore, the model in previous row is taken.}
\label{tab:aic}
\end{table*}

\subsection{Comparison with high-angular-resolution data}
\label{sec:pv}
The multi-component fitting reveals several kinematic components. Though the best-fit model has been checked (Figure \ref{fig:fitting}), here we further evaluate the fitting results with high-angular-resolution data. We compare the central velocity, linewidth, and source size from the best-fit with the high-angular-resolution data.
aGg'-\ce{(CH2OH)2}, \ce{CH3CHO}, and \ce{CH3OCHO} emission was detected with ALMA at an angular resolution of $\sim0\farcs20\times0\farcs11$ and PA=-0.9$^\circ$ ($\sim30-60$ au) \citep{di22}.

Figure \ref{fig:pv} shows the position-velocity (PV) diagrams from two selected transitions of aGg'-\ce{(CH2OH)2}, \ce{CH3CHO}, and \ce{CH3OCHO} in the high resolution data, for which the PV cut is taken along the east-west direction through VLA4A and VLA4B (Figure \ref{fig:mom0}). Here we compare the best-fit $V_{\rm LSR}$, $\Delta V$, and sizes with the PV diagrams.
In the PV-diagram, we plot an ellipse representing each kinematic component from the best-fit model. The width, height, and x-axis location correspond to the $\Delta V$, source sizes, and $V_{\rm LSR}$. The y-axis offsets of ellipses are manually adjusted by eyes. The height of the ellipse is given by $({\rm size}^2+{\rm beam_{EW}}^2)^{1/2}$ where ${\rm beam_{EW}}$ is the the beam size along the East West PV cut.
While the $V_{\rm LSR}$ and $\Delta V$ can be directly seen in the spectral profiles (Figure \ref{fig:fitting}), the source size is obtained based on the line intensity ratios from multiple transitions with their opacities $\tau$.
In other words, an accurate source size suggests a good $\tau$ correction with a reasonable column density as well as the temperature. 
As a result, each component from the lower-spatial-resolution NOEMA data is a relatively good match of the components seen in the high-spatial-resolution ALMA data, suggesting reasonable fitting of the NOEMA data at least for the three COMs, aGg'-\ce{(CH2OH)2}, \ce{CH3CHO}, and \ce{CH3OCHO}.
Again, the broad-line components in aGg'-\ce{(CH2OH)2} and \ce{CH3CHO} can be ambiguous; the best-fit source sizes for the broad-line component, $0\farcs034$ and $0\farcs083$, are smaller than the beam of the ALMA observation.
In such a case, the ALMA data do not resolve these components and only provide an upper limit to their source size.
For the other components, derived source sizes are also on the same order or less than the beam of the ALMA observations.

\section{Discussion}
\label{sec:dis}
\subsection{Sizes of the O-bearing COM emission}
To probe the origin of COM emission, it is crucial to estimate the size of the emitting area.
The origin of COM emission can be roughly categorized into three groups (1) hot corinos, (2) outflows, and (3) accretion shock/disk atmosphere by \citet{be20}.

With the uv-domain Gaussian fitting, we find that the COM emission is point-source like with our NOEMA observations at an angular resolution of $1\farcs2\times0\farcs7$ ($350\times200$ au). This likely rules out the scenario that trace the outflow cavity walls which is usually extended.

Expected from the bolometric luminosity of SVS13A 44 $L_\odot$, \citet{be20} estimate a radius of $>40$ au where the temperature is about the sublimation temperature of COMs $\sim$100-150 K.
Because this size is similar to the size of COM emission (see also \citealt{hs19,hs23}), \citet{be20} classify the COM emission in SVS13A as a hot corino. 

\begin{figure}
\centering
\includegraphics[width=0.5\textwidth]{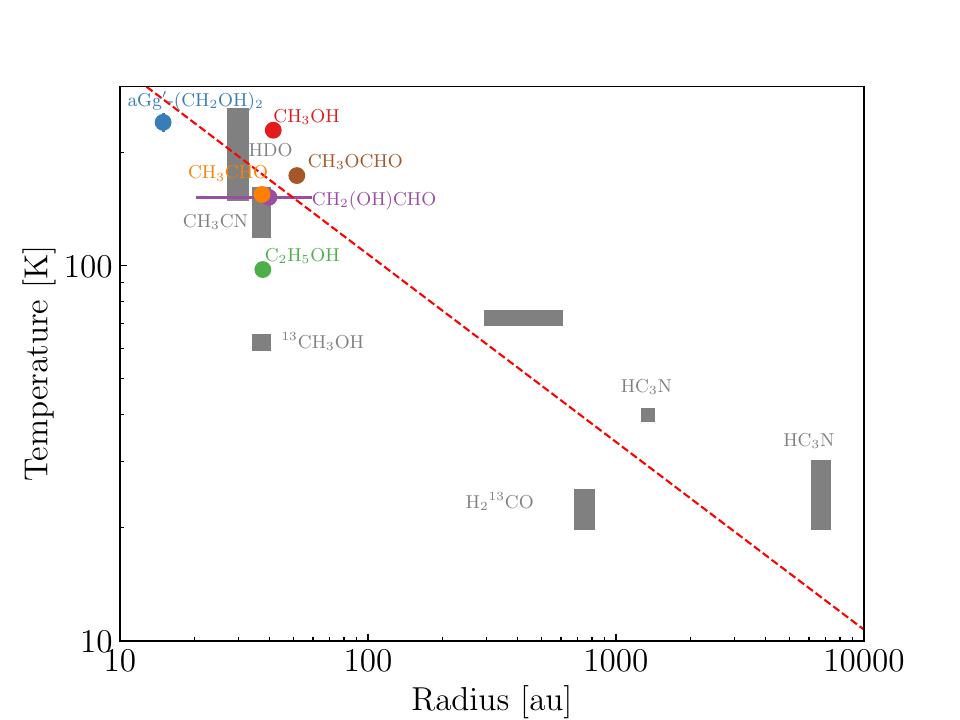}
\caption{
Figure 9 from \citet{bi22a} with data points from COMs in this work.
}
\label{fig:r_vs_t}
\end{figure}

With the LTE model of one Gaussian component (Table \ref{tab:mcmc}), we derive a FWHM size of each COM emission: $0\farcs11$ (31 au for aGg'-\ce{(CH2OH)2}), $0\farcs25$ (74 au for \ce{CH3CHO}), $0\farcs25$ (74 au for \ce{C2H5OH}), $0\farcs22$ (65 au for \ce{CH3OCHO}), and $0\farcs31$ (92 au for \ce{CH2(OH)CHO}). These sizes are derived based on a number of transitions covering optically thin to thick emission (except for \ce{CH2(OH)CHO}, see Appendix \ref{app:fitting}), providing an estimate of source size independent of the angular resolution of the observation (see also Sect.\ \ref{sec:pv}). These source sizes are within the sublimation radius ($\sim$$30-80$ au for $150-100$ K) estimated by \citet{be20}, but different O-bearing COMs trace different sizes ($r=16-45$ au).
Intuitively, one might speculate that it is due to different sublimation temperatures of COMs or different chemical processes at work in different locations of the source; a trend of the decreasing temperature with increasing radius from $\sim$30$-$8000 au is found using the derived source sizes and temperatures from molecules (Figure 9 in \citealt{bi22a}). 
The derived $T_{\rm ex}$ with source sizes (radius) are consistent with the correlation found in \citet{bi22a}.
Toward the small scale with COM emission, we do not seen clear trend with the limited number of data points (Figure \ref{fig:r_vs_t}).

\subsection{Complex structures}
\subsubsection{Kinematics}
The six O-bearing COMs show different profiles (Figure \ref{fig:stack}), suggesting that they come from different regions within the SVS13A system. Although the emission is not spatially resolved, their Gaussian centers at different velocities further support this scenario (Figure \ref{fig:uv_cen}).

Among the six O-bearing COMs, we found velocity gradients from the west to the east in \ce{CH3OH}, \ce{C2H5OH}, \ce{CH3CHO}, \ce{CH3OCHO}, while aGg'-\ce{(CH2OH)2} and \ce{CH2(OH)CHO} emission do not show a velocity gradient. 
For aGg'-\ce{(CH2OH)2}, it is consistent with the ALMA high-resolution observations by \citet{di22} in which the aGg'-\ce{(CH2OH)2} emission only traces the VLA4A disk.
The double-peak line profiles around the $V_{\rm LSR}$ of VLA4A (Figure \ref{fig:stack}) from aGg\'-\ce{(CH2OH)2} and \ce{CH2(OH)CHO} also imply that their emission is dominated by a rotating disk/inner envelope of VLA4A.
For the other four COMs, although the velocity gradients share a similar direction from the west to the east, they have indeed different slopes. The fitted Gaussian centers may be considered as a flux-weighted centroid in a case of unresolved extended emission at the given velocity. Therefore, together with the multiple-peak spectral profiles (Figure \ref{fig:stack}), we speculate that the emission of these four COMs traces not only the VLA4A disk but also the VLA4B disk and/or the circumbinary disk although the VLA4A disk likely dominates it based on both the velocity and spatial information.
Even for \ce{CH3OCHO} with the most extended structure (Figure \ref{fig:uv_cen}), the red-shifted emission near the $V_{\rm LSR}=9.33$ km s$^{-1}$ is located around the mid-point of VLA4A and VLA4B. This is also consistent with the high-angular-resolution ALMA data (Figure \ref{fig:pv}).

COM emission with different optical depths can trace different layers, as we found for \ce{CH3{}^{13}CN} and \ce{CH3CN} \citep{hs23}. 
In case of protostellar envelopes, intuitively, the optically thick lines of COMs should trace the outer region while the thin transitions should trace the inner dense region.
It is more complicated in Fig. \ref{fig:uv_cen} as each velocity step has a different optical depth. 
Thus, we should see the spatial distribution from low to high velocity goes across from optically thin (wing) to optically thick (middle) and back to optically thin (wing). In this case, in additional to the optically thin transitions, the blue/red-shifted line wing should point toward the center.
However, this is unlikely the case; the most optically thick line of \ce{CH3OH} ($\tau_{\rm 1comp.}=7.3$, Table \ref{tab:tran_ch3oh}), likely traces the inner region where a large velocity gradient is present (Figure \ref{fig:uv_cen}), while the \ce{CH3OCHO} emission is associated with the outer region with $\tau_{\rm 1comp.}=0.6-1.7$ (Table \ref{tab:tran_ch3ocho}).
Alternatively, if the COM emission comes from a disk, it is possible the optically thick emission lines trace the inner part of the disk where the density and temperature are high.
However, as mentioned in section \ref{sec:gau}, the spatial distributions are similar from different transitions with different optical depths for each COM (Figure \ref{fig:uv_cen}).
This suggests that the discrepancy seen in the COM distribution most likely originate from chemical, rather than optical depth effects.

\subsubsection{Physical conditions}
Traditionally, to derive physical conditions of the gas traced by COMs, rotation diagrams and/or simple one-component Gaussian models are used. It is however frequently seen in rotation diagrams that multiple components exist with different temperatures (slopes) when many transitions cover wide ranges of energy levels. 
With this in mind, to robustly derive the physical conditions, we need to first identify the kinematic components with high-angular and spectral resolution data with as many transitions as possible. However, this is usually very time expensive in observations.

The abundance ratios relative to \ce{CH3OH} or other species, e.g. [X]/\ce{CH3OH}, have been commonly used to study the chemical composition (e.g., \citealt{ca18,be20,jo20,ya21}). 
The abundance ratios ($\frac{[X]}{[\ce{CH3OH}]}$ from one-velocity model, Table \ref{tab:mcmc}) toward SVS13A are slightly different by a factor of 0.5-2.6 to the column density ratios derived by \citet{be20} except for aGg'-\ce{(CH2OH)2} with a factor of 4.
In \citet{be20}, the column density is derived based on the rotation diagram under an assumption of source size of $0\farcs3$ (Table 5 in the paper). For aGg'-\ce{(CH2OH2}, the difference is likely caused by a very different rotational temperature 98 K from \citet{be20} and 240 K in this work. Here, we have included transitions with $E_{\rm up}\gtrsim$300 K as \citet{be20} use transitions with $E_{\rm up}<250$ K (Fig. E.29 in the paper). In the multiple-velocity component fitting, we find a possible component with a rotation temperature of 470 K that is dominant among the molecule. This explains the difference in the abundance ratio of aGg'-\ce{(CH2OH)2}, i.e., $\frac{[\rm aGg'-\ce{(CH2OH)2}]}{[\ce{CH3OH}]}$, derived from \citet{be20} and this work.
However, toward SVS13A, it is clear that COMs are associated with different structures. Unless the COM emission is both spectrally and spatially resolved, the discussion of abundance ratios can mix together contributions of distinct structures, providing no quantitative physical constraints needed for chemical models.

\begin{figure}
\includegraphics[width=0.5\textwidth]{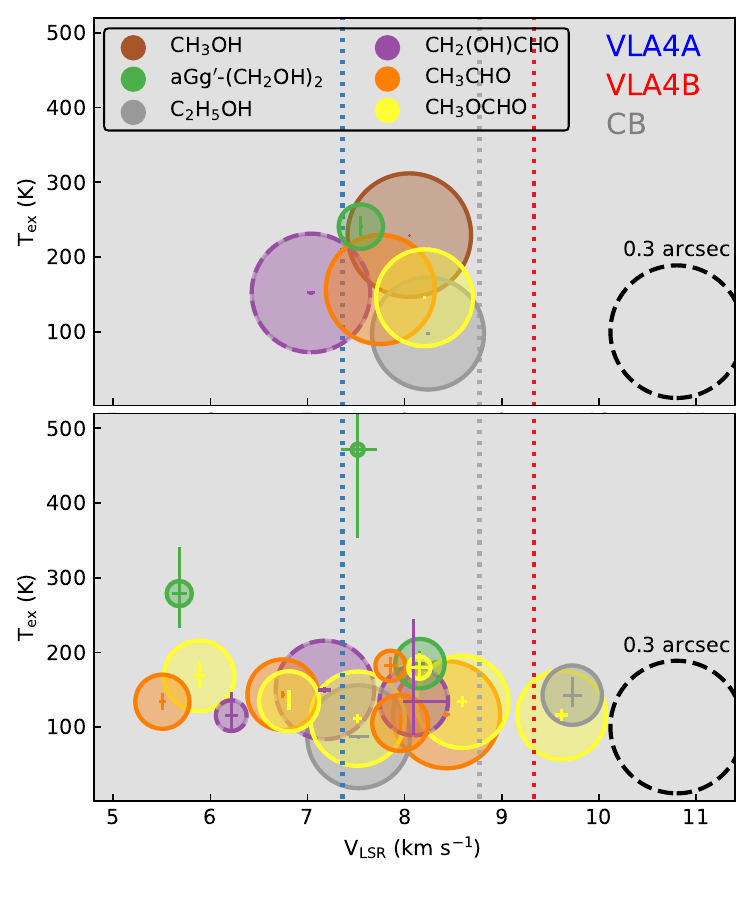}
\caption{Excitation temperature as a function of central velocity for the six O-bearing COMs. The top panel shows the results from single component fitting while the bottom panel shows the results from the multi-component fit. The size of the circle represents the source size; \ce{CH2(OH)CHO} circle has dashed outline as the source size of \ce{CH2(OH)CHO} can be degenerate with column density due to lack of optically-thick transition.
The dashed circle at the bottom right corner indicate the size of 0.3 arcsec.
}
\label{fig:com_mcmc}
\end{figure}

Our NOEMA observations cover a number of transitions for modeling at a high-spectral resolution (\Cref{tab:tran_ch3oh,tab:tran_ch2oh2,tab:tran_c2h5oh,tab:tran_ch2ohcho,tab:tran_ch3cho,tab:tran_ch3ocho}). Thus, multiple kinematic components can be disentangled and their physical conditions can be derived (Table \ref{tab:mcmc}, see also \citealt{hs23} for \ce{CH3CN}).
Figure \ref{fig:com_mcmc} shows the temperature of these kinematic components as a function of velocity (Table \ref{tab:mcmc}). 
For the one-component model, it unveils that the different O-bearing COMs are tracing different regions with different temperatures and source sizes.
The multi-component fitting further unveils several components with temperatures in a range from a few tens to a few hundreds K.
It is likely that the one-component model usually gives a temperature between those derived from multiple-component model. 
However, the total amount of molecules (i.e., $N_{\rm tot}\times$area in Table \ref{tab:mcmc}) is always higher in multiple-component fitting; the sum of the $N_{\rm tot}\times$area from the multiple-component fitting is larger than that of single component fitting by a factor of $1.1-1.5$ (see $\frac{[X]}{[\ce{CH3OH}]}$ in Table \ref{tab:mcmc}. That is because the multiple-component fitting process may unveil hidden components from the complex line profile \citep{hs23}. Also, these components are assumed to be spatially separated so that the integrated $\tau$ from the multi-velocity model can be larger than that of the single component.

\subsection{Origin of COM emission in SVS13A}
Thanks to the capabilities of the NOEMA PolyFiX correlator, we have identified multiple velocity components toward SVS13A, and better quantified the kinematics and physical conditions using six O-bearing COMs.
It is clear that the COM emission traces different kinematic components toward SVS13A with different physical conditions (temperature, column density and size). 
This suggests that the COM emission in SVS13A originates from a complex structure, which may not be explained by a unique scenario.

It is interesting to compare with the well-studied hot corino protobinary system IRAS16293-2422A (hereafter IRAS16293A) at an earlier evolutionary stage than SVS13A \citep{pi12,jo16,ce22}. With high-angular resolution ALMA observations, \citet{ma20} found complex structures toward IRAS16293A using high excitation transitions of \ce{HNCO}, \ce{NH2CHO}, and \ce{t-HCOOH}. 
\citet{ma22} considered that these complex structures are associated with shocked gas or a gravitationally unstable disk (spiral). This result is consistent with the inhomogeneous dust/gas temperature \citep{va20} (see also \citealt{za21}).
It is proposed that shocked gas might respond to local temperature increases with sublimation of COMs \citep{oy16,ma22}. 
In addition, an infalling streamer inducing local shocks has been found by \citet{ga22} via SO toward a more evolved young stellar object HL Tau (see also \citealt{va21}); an infalling streamer is also suggested toward IRAS16293A by \citet{mu22}. 
This hints at a scenario where the physical and chemical properties of an inner region ($\lesssim300$ au) is affected by large-scale ($\sim$a few thousand au) infalling streamers (see \citealt{pi22} and references therein).


Toward SVS13A, \citet{hs23} suggest that the continuum spiral structure connecting to VLA4A is associated with a streamer from larger scales which is possibly infalling. \citet{bi22b} speculate that streamer-induced shocks sputter icy COMs from dust grains given the asymmetric structure of COM emission toward SVS13A from high-angular resolution data.
In this work, we find that the COM emission traces complex structures, so it is unlikely only due to heating from the central protostellar source. 
Thus, we consider that shocks from materials infalling onto and impacting circumstellar or circumbinary disks play an important role in feeding COMs from dust grains, leading to the complex structure in COM emission. 
Other possibilities are shocks produced by the self-gravity or tidal forces within the disks themselves.
To understand this requires high-angular resolution observations, coupled with high-spectral resolution and sensitivity.

\section{Conclusion}
\label{sec:con}
We present NOEMA observations toward SVS13A with six selected O-bearing COMs. The spectral profiles reveal multiple kinematic velocity components tracing different physical entities as well as their distinctive spatial distributions, suggesting that the emission from each COM species can come from different regions. We derive their physical conditions, supporting that these COMs trace a complex structure in the protobinary system SVS13A.
Our main conclusions are summarized as follows:

\begin{enumerate}
\item Toward the protobinary system SVS13A, we find that for a given COM species, the spatial distributions of different transitions show similar structures based on the uv-space Gaussian fitting at different velocities.
However, different COMs do not trace the same kinematic components.
This causes a caveat when discussing the abundance ratios of COMs as well as chemical properties without kinematically resolving the systems. 
The measurement of relative COM abundances requires high spectral (as well as angular) resolution observations.

\item By covering numerous transitions with high-spectral resolution, we decompose the emission of each COM into several kinematic components. 
We suggest that the traditional rotational diagram or one-component fitting can underestimate the total amount of molecules.
We found that each COM can contain multiple velocity components within 10s of au radius in the hot corino, tracing different physical conditions.

\item We found that the COM emission from SVS13A is associated with an inhomogeneous complex structure. The small source sizes show that the emission unlikely arise from outflow cavity walls.
This complex structure is unlikely caused only by heating of the central protostar. We conclude the COM emission most likely comes from shocked gas at the disk scale. A possibility to induce the shocked gas is the large-scale streamer \citet{hs23}, which may be infalling directly onto the disk and produce inhomogenous shocked regions.

\end{enumerate}
Our results suggest that different COMs can trace very different kinematic components in a system. This causes a caveat in comparing the abundances before resolving it. The emission of an individual COM can be associated with multiple components. This suggests that COM emission is influenced by localized shock activities instead of only the protostellar heating. 

\begin{acknowledgements}
We are grateful for the anonymous referee for the thorough and insightful comments that helped to improve this paper significantly.
T.-H. H., J.E.P., P.C., M.T.V, and M.J.M.
acknowledge the support by the Max Planck Society.
The authors thank Eleonora Bianchi Valerio Lattanzi, Silvia Spezzano, and Christian Endres for the chemical modeling and CDMS.
D. S.-C. is supported by an NSF Astronomy and Astrophysics Postdoctoral Fellowship under award AST-2102405.
This work is based on observations carried out under project number L19MB with the IRAM NOEMA Interferometer. IRAM is supported by INSU/CNRS (France), MPG (Germany) and IGN (Spain).
This paper makes use of the following ALMA data: ADS/JAO.ALMA\#2013.1.00031.S and 2016.1.01305. ALMA is a partnership of ESO (representing its member states), NSF (USA) and NINS (Japan), together with NRC (Canada), MOST and ASIAA (Taiwan), and KASI (Republic of Korea), in cooperation with the Republic of Chile. The Joint ALMA Observatory is operated by ESO, AUI/NRAO and NAOJ.
SM is supported by a Royal Society University Research Fellowship (URF-R1-221669).
D.S. and Th. H. acknowledge support from the European Research Council under the Horizon 2020 Framework Program via the ERC Advanced Grant Origins 83 24 28.
AF thanks the Spanish MICIN for funding support from PID2019-106235GB-I00 and the European Research Council (ERC) for funding under the Advanced Grant project SUL4LIFE, grant agreement No101096293.
\end{acknowledgements}

\bibliographystyle{aa}

\begin{thebibliography}{}
\bibitem[Alves et al.(2020)]{al20} Alves, F.~O., Cleeves, L.~I., Girart, J.~M., et al.\ 2020, \apjl, 904, L6. doi:10.3847/2041-8213/abc550
\bibitem[Anglada et al.(2004)]{an04} Anglada, G., Rodr{\'\i}guez, L.~F., Osorio, M., et al.\ 2004, \apjl, 605, L137. doi:10.1086/420782
\bibitem[Arce et al.(2008)]{ar08} Arce, H.~G., Santiago-Garc{\'\i}a, J., J{\o}rgensen, J.~K., et al.\ 2008, \apjl, 681, L21. doi:10.1086/590110
\bibitem[Astropy Collaboration et al.(2013)]{as13} Astropy Collaboration, Robitaille, T.~P., Tollerud, E.~J., et al.\ 2013, \aap, 558, A33. doi:10.1051/0004-6361/201322068
\bibitem[Bacmann et al.(2012)]{ba12} Bacmann, A., Taquet, V., Faure, A., et al.\ 2012, \aap, 541, L12. doi:10.1051/0004-6361/201219207
\bibitem[Belloche et al.(2020)]{be20} Belloche, A., Maury, A.~J., Maret, S., et al.\ 2020, \aap, 635, A198. doi:10.1051/0004-6361/201937352
\bibitem[Bianchi et al.(2017)]{bi17} Bianchi, E., Codella, C., Ceccarelli, C., et al.\ 2017, \mnras, 467, 3011. doi:10.1093/mnras/stx252
\bibitem[Bianchi et al.(2019)]{bi19} Bianchi, E., Codella, C., Ceccarelli, C., et al.\ 2019, \mnras, 483, 1850. doi:10.1093/mnras/sty2915
\bibitem[Bianchi et al.(2022a)]{bi22a} Bianchi, E., Ceccarelli, C., Codella, C., et al.\ 2022a, \aap, 662, A103. doi:10.1051/0004-6361/202141893
\bibitem[Bianchi et al.(2022b)]{bi22b} Bianchi, E., L{\'o}pez-Sepulcre, A., Ceccarelli, C., et al.\ 2022b, \apjl, 928, L3. doi:10.3847/2041-8213/ac5a56
\bibitem[Blake et al.(1987)]{bl87} Blake, G.~A., Sutton, E.~C., Masson, C.~R., et al.\ 1987, \apj, 315, 621. doi:10.1086/165165
\bibitem[Bottinelli et al.(2004)]{bo04} Bottinelli, S., Ceccarelli, C., Lefloch, B., et al.\ 2004, \apj, 615, 354. doi:10.1086/423952
\bibitem[Bottinelli et al.(2007)]{bo07} Bottinelli, S., Ceccarelli, C., Williams, J.~P., et al.\ 2007, \aap, 463, 601. doi:10.1051/0004-6361:20065139
\bibitem[Bouvier et al.(2021)]{bo21} Bouvier, M., L{\'o}pez-Sepulcre, A., Ceccarelli, C., et al.\ 2021, \aap, 653, A117. doi:10.1051/0004-6361/202141157
\bibitem[Cabedo et al.(2021)]{ca21} Cabedo, V., Maury, A., Girart, J.~M., et al.\ 2021, \aap, 653, A166. doi:10.1051/0004-6361/202140754
\bibitem[Calcutt et al.(2018)]{ca18} Calcutt, H., J{\o}rgensen, J.~K., M{\"u}ller, H.~S.~P., et al.\ 2018, \aap, 616, A90. doi:10.1051/0004-6361/201732289
\bibitem[Caselli \& Ceccarelli(2012)]{ca12} Caselli, P. \& Ceccarelli, C.\ 2012, \aapr, 20, 56. doi:10.1007/s00159-012-0056-x
\bibitem[Cazaux et al.(2003)]{ca03} Cazaux, S., Tielens, A.~G.~G.~M., Ceccarelli, C., et al.\ 2003, \apjl, 593, L51. doi:10.1086/378038
\bibitem[Ceccarelli et al.(2017)]{ce17} Ceccarelli, C., Caselli, P., Fontani, F., et al.\ 2017, \apj, 850, 176. doi:10.3847/1538-4357/aa961d
\bibitem[Ceccarelli et al.(2022)]{ce22} Ceccarelli, C., Codella, C., Balucani, N., et al.\ 2022, arXiv:2206.13270
\bibitem[Choudhury et al.(2020)]{ch20} Choudhury, S., Pineda, J.~E., Caselli, P., et al.\ 2020, \aap, 640, L6. doi:10.1051/0004-6361/202037955
\bibitem[Christen et al.(2001)]{ch01} Christen, D., Coudert, L.~H., Larsson, J.~A., et al.\ 2001, Journal of Molecular Spectroscopy, 205, 185. doi:10.1006/jmsp.2000.8263
\bibitem[Codella et al.(2018)]{co18} Codella, C., Bianchi, E., Tabone, B., et al.\ 2018, \aap, 617, A10. doi:10.1051/0004-6361/201832765
\bibitem[Codella et al.(2021)]{co21} Codella, C., Bianchi, E., Podio, L., et al.\ 2021, \aap, 654, A52. doi:10.1051/0004-6361/202141485
\bibitem[Coletta et al.(2020)]{co20} Coletta, A., Fontani, F., Rivilla, V.~M., et al.\ 2020, \aap, 641, A54. doi:10.1051/0004-6361/202038212
\bibitem[Csengeri et al.(2018)]{cs18} Csengeri, T., Bontemps, S., Wyrowski, F., et al.\ 2018, \aap, 617, A89. doi:10.1051/0004-6361/201832753
\bibitem[Csengeri et al.(2019)]{cs19} Csengeri, T., Belloche, A., Bontemps, S., et al.\ 2019, \aap, 632, A57. doi:10.1051/0004-6361/201935226
\bibitem[Diaz-Rodriguez et al.(2022)]{di22} Diaz-Rodriguez, A.~K., Anglada, G., Bl{\'a}zquez-Calero, G., et al.\ 2022, \apj, 930, 91. doi:10.3847/1538-4357/ac3b50
\bibitem[Drozdovskaya et al.(2015)]{dr15} Drozdovskaya, M.~N., Walsh, C., Visser, R., et al.\ 2015, \mnras, 451, 3836. doi:10.1093/mnras/stv1177
\bibitem[Endres et al.(2016)]{en16} Endres, C.~P., Schlemmer, S., Schilke, P., et al.\ 2016, Journal of Molecular Spectroscopy, 327, 95. doi:10.1016/j.jms.2016.03.005
\bibitem[Foreman-Mackey et al.(2013)]{fo13} Foreman-Mackey, D., Hogg, D.~W., Lang, D., et al.\ 2013, \pasp, 125, 306. doi:10.1086/670067
\bibitem[Garufi et al.(2022)]{ga22} Garufi, A., Dominik, C., Ginski, C., et al.\ 2022, \aap, 658, A137. doi:10.1051/0004-6361/202141692
\bibitem[Gieser et al.(2023)]{gi23} Gieser, C., Beuther, H., Semenov, D., et al.\ 2023, \aap, 674, A160. doi:10.1051/0004-6361/202245249
\bibitem[Ginski et al.(2021)]{gi21} Ginski, C., Facchini, S., Huang, J., et al.\ 2021, \apjl, 908, L25. doi:10.3847/2041-8213/abdf57
\bibitem[Goodman \& Weare(2010)]{go10} Goodman, J. \& Weare, J.\ 2010, Communications in Applied Mathematics and Computational Science, 5, 65. doi:10.2140/camcos.2010.5.65
\bibitem[Harsono et al.(2013)]{ha13} Harsono, D., Visser, R., Bruderer, S., et al.\ 2013, \aap, 555, A45. doi:10.1051/0004-6361/201220885
\bibitem[Herbst \& van Dishoeck(2009)]{he09} Herbst, E. \& van Dishoeck, E.~F.\ 2009, \araa, 47, 427. doi:10.1146/annurev-astro-082708-101654
\bibitem[Hogg \& Foreman-Mackey(2018)]{ho18} Hogg, D.~W. \& Foreman-Mackey, D.\ 2018, \apjs, 236, 11. doi:10.3847/1538-4365/aab76e
\bibitem[Hsieh et al.(2019)]{hs19} Hsieh, T.-H., Murillo, N.~M., Belloche, A., et al.\ 2019, \apj, 884, 149
\bibitem[Hsieh et al.(2023)]{hs23} Hsieh, T.-H., Segura-Cox, D.~M., Pineda, J.~E., et al.\ 2023, \aap, 669, A137. doi:10.1051/0004-6361/202244183
\bibitem[Hunter (2007)]{matplotlib}Hunter, J. D. 2007, Computing in Science and Engineering, 9, 90
\bibitem[J{\o}rgensen et al.(2004)]{jo04} J{\o}rgensen, J.~K., Hogerheijde, M.~R., Blake, G.~A., et al.\ 2004, \aap, 415, 1021. doi:10.1051/0004-6361:20034216
\bibitem[J{\o}rgensen et al.(2016)]{jo16} J{\o}rgensen, J.~K., van der Wiel, M.~H.~D., Coutens, A., et al.\ 2016, \aap, 595, A117. doi:10.1051/0004-6361/201628648
\bibitem[J{\o}rgensen et al.(2018)]{jo18} J{\o}rgensen, J.~K., M{\"u}ller, H.~S.~P., Calcutt, H., et al.\ 2018, \aap, 620, A170. doi:10.1051/0004-6361/201731667
\bibitem[J{\o}rgensen et al.(2020)]{jo20} J{\o}rgensen, J.~K., Belloche, A., \& Garrod, R.~T.\ 2020, \araa, 58, 727. doi:10.1146/annurev-astro-032620-021927
\bibitem[Lef{\`e}vre et al.(2017)]{le17} Lef{\`e}vre, C., Cabrit, S., Maury, A.~J., et al.\ 2017, \aap, 604, L1. doi:10.1051/0004-6361/201730766
\bibitem[Lefloch et al.(2018)]{le18} Lefloch, B., Bachiller, R., Ceccarelli, C., et al.\ 2018, \mnras, 477, 4792. doi:10.1093/mnras/sty937
\bibitem[Maureira et al.(2020)]{ma20} Maureira, M.~J., Pineda, J.~E., Segura-Cox, D.~M., et al.\ 2020, \apj, 897, 59. doi:10.3847/1538-4357/ab960b
\bibitem[Maureira et al.(2022)]{ma22} Maureira, M.~J., Gong, M., Pineda, J.~E., et al.\ 2022, \apjl, 941, L23. doi:10.3847/2041-8213/aca53a
\bibitem[McMullin et al.(2007)]{casa} McMullin, J.~P., Waters, B., Schiebel, D., et al.\ 2007, Astronomical Data Analysis Software and Systems XVI, 376, 127
\bibitem[M{\"o}ller et al.(2017)]{mo17} M{\"o}ller, T., Endres, C., \& Schilke, P.\ 2017, \aap, 598, A7. doi:10.1051/0004-6361/201527203
\bibitem[M{\"u}ller et al.(2005)]{mu05} M{\"u}ller, H.~S.~P., Schl{\"o}der, F., Stutzki, J., et al.\ 2005, Journal of Molecular Structure, 742, 215. doi:10.1016/j.molstruc.2005.01.027
\bibitem[Murillo et al.(2022)]{mu22} Murillo, N.~M., van Dishoeck, E.~F., Hacar, A., et al.\ 2022, \aap, 658, A53. doi:10.1051/0004-6361/202141250
\bibitem[Ortiz-Le{\'o}n et al.(2018)]{or18} Ortiz-Le{\'o}n, G.~N., Loinard, L., Dzib, S.~A., et al.\ 2018, \apjl, 869, L33. doi:10.3847/2041-8213/aaf6ad
\bibitem[Oya et al.(2016)]{oy16} Oya, Y., Sakai, N., L{\'o}pez-Sepulcre, A., et al.\ 2016, \apj, 824, 88. doi:10.3847/0004-637X/824/2/88
\bibitem[Palau et al.(2017)]{pa17} Palau, A., Walsh, C., S{\'a}nchez-Monge, {\'A}., et al.\ 2017, \mnras, 467, 2723. doi:10.1093/mnras/stx004
\bibitem[Pineda et al.(2012)]{pi12} Pineda, J.~E., Maury, A.~J., Fuller, G.~A., et al.\ 2012, \aap, 544, L7. doi:10.1051/0004-6361/201219589
\bibitem[Pineda et al.(2020)]{pi20} Pineda, J.~E., Segura-Cox, D., Caselli, P., et al.\ 2020, Nature Astronomy, 4, 1158. doi:10.1038/s41550-020-1150-z
\bibitem[Pineda et al.(2022)]{pi22} Pineda, J.~E., Arzoumanian, D., Andr{\'e}, P., et al.\ 2022, arXiv:2205.03935
\bibitem[Robitaille \& Bressert(2012)]{ro12} Robitaille, T. \& Bressert, E.\ 2012, Astrophysics Source Code Library. ascl:1208.017
\bibitem[Sargent \& Beckwith(1987)]{sa87} Sargent, A.~I. \& Beckwith, S.\ 1987, \apj, 323, 294. doi:10.1086/165827
\bibitem[Segura-Cox et al.(2018)]{se18} Segura-Cox, D.~M., Looney, L.~W., Tobin, J.~J., et al.\ 2018, \apj, 866, 161. doi:10.3847/1538-4357/aaddf3
\bibitem[Spezzano et al.(2016)]{sp16} Spezzano, S., Bizzocchi, L., Caselli, P., et al.\ 2016, \aap, 592, L11. doi:10.1051/0004-6361/201628652
\bibitem[Sugimura et al.(2011)]{su11} Sugimura, M., Yamaguchi, T., Sakai, T., et al.\ 2011, \pasj, 63, 459. doi:10.1093/pasj/63.2.459
\bibitem[Tafalla et al.(2006)]{ta06} Tafalla, M., Santiago-Garc{\'\i}a, J., Myers, P.~C., et al.\ 2006, \aap, 455, 577. doi:10.1051/0004-6361:20065311
\bibitem[Thieme et al.(2022)]{th22} Thieme, T.~J., Lai, S.-P., Lin, S.-J., et al.\ 2022, \apj, 925, 32. doi:10.3847/1538-4357/ac382b
\bibitem[Tobin et al.(2016)]{to16} Tobin, J.~J., Looney, L.~W., Li, Z.-Y., et al.\ 2016, \apj, 818, 73. doi:10.3847/0004-637X/818/1/73
\bibitem[Tobin et al.(2018)]{to18} Tobin, J.~J., Looney, L.~W., Li, Z.-Y., et al.\ 2018, \apj, 867, 43. doi:10.3847/1538-4357/aae1f7
\bibitem[Tychoniec et al.(2020)]{ty20} Tychoniec, {\L}., Manara, C.~F., Rosotti, G.~P., et al.\ 2020, \aap, 640, A19. doi:10.1051/0004-6361/202037851
\bibitem[Valdivia-Mena et al.(2022)]{va22} Valdivia-Mena, M.~T., Pineda, J. E., Segura-Cox, D M., et al.\ 2022, \aap, submitted
\bibitem[Van Der Walt et al. (2011)]{numpy}Van Der Walt, S., Colbert, S. C., \& Varoquaux, G. 2011, ArXiv e-prints, arXiv:1102.1523 [cs.MS]
\bibitem[van Dishoeck et al.(1995)]{va95} van Dishoeck, E.~F., Blake, G.~A., Jansen, D.~J., et al.\ 1995, \apj, 447, 760. doi:10.1086/175915
\bibitem[van 't Hoff et al.(2020)]{va20} van 't Hoff, M.~L.~R., van Dishoeck, E.~F., J{\o}rgensen, J.~K., et al.\ 2020, \aap, 633, A7. doi:10.1051/0004-6361/201936839
\bibitem[Vastel et al.(2022)]{vas22} Vastel, C., Alves, F., Ceccarelli, C., et al.\ 2022, \aap, 664, A171. doi:10.1051/0004-6361/202243414
\bibitem[van Gelder et al.(2021)]{va21} van Gelder, M.~L., Tabone, B., van Dishoeck, E.~F., et al.\ 2021, \aap, 653, A159. doi:10.1051/0004-6361/202141591
\bibitem[Virtanen et al. (2020)]{scipy}Virtanen, P., Gommers, R., Oliphant, et al. (2020). SciPy 1.0: Fundamental Algorithms for Scientific Computing in Python. Nature Methods.
\bibitem[van Dishoeck(2014)]{va14} van Dishoeck, E.~F.\ 2014, Faraday Discussions, 168, 9. doi:10.1039/C4FD00140K
\bibitem[Yang et al.(2021)]{ya21} Yang, Y.-L., Sakai, N., Zhang, Y., et al.\ 2021, \apj, 910, 20. doi:10.3847/1538-4357/abdfd6
\bibitem[Zamponi et al.(2021)]{za21} Zamponi, J., Maureira, M.~J., Zhao, B., et al.\ 2021, \mnras, 508, 2583. doi:10.1093/mnras/stab2657
\end{thebibliography}


{\it Software:} Numpy (\citealt{numpy}),
Scipy (\citealt{scipy}), 
APLpy (\citealt{ro12}),
Matplotlib (\citealt{matplotlib}), 
Astropy (\citealt{as13})
CASA (\citealt{casa})

\appendix
\setcounter{figure}{0}
\setcounter{table}{0}
\counterwithin{figure}{section}
\counterwithin{equation}{section}
\counterwithin{table}{section}

\section{Line stacking}
\label{app:stack}
\begin{figure*}
\centering
\includegraphics[width=0.95\textheight,angle=-90]{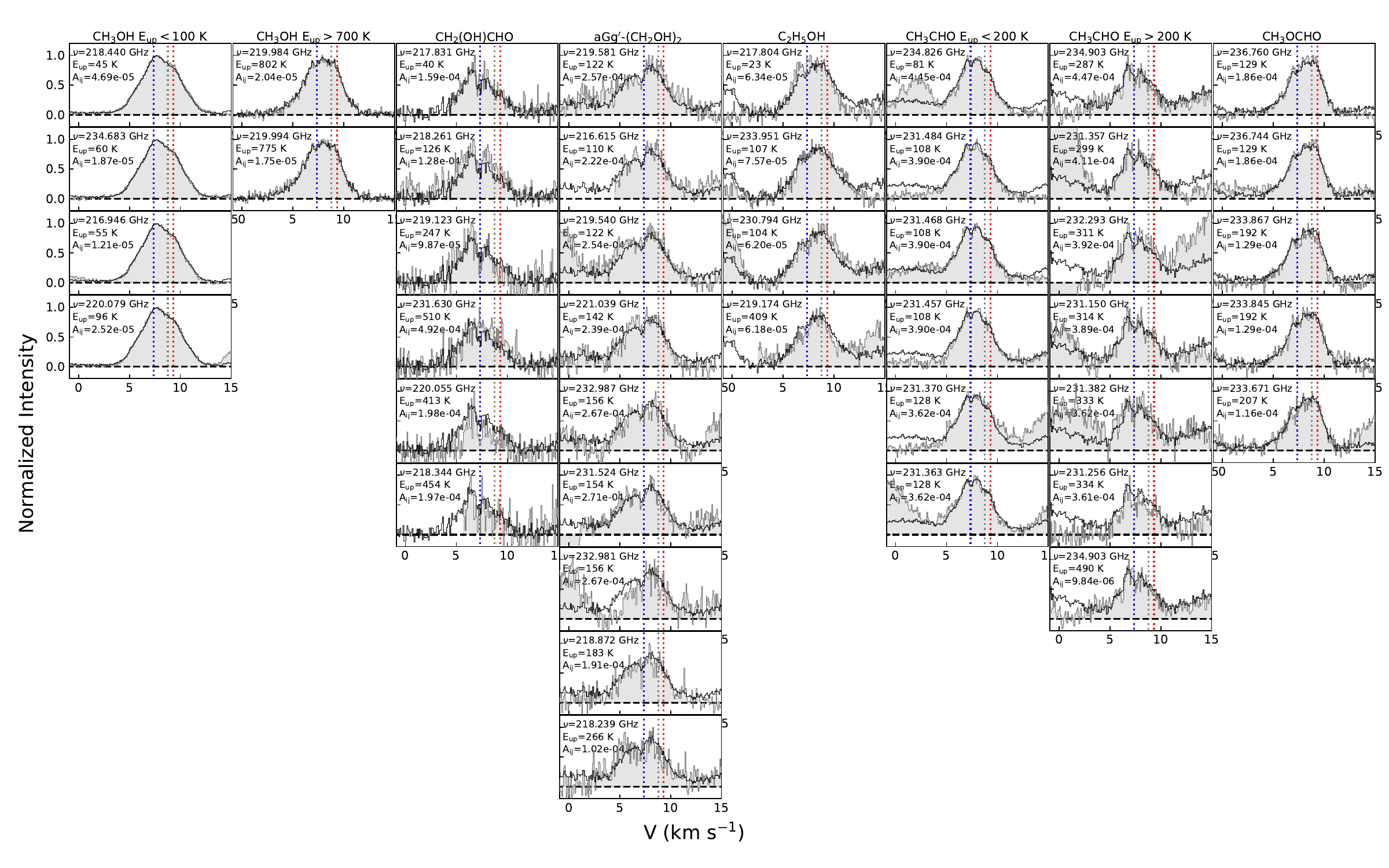}
\caption{All spectral profiles used for the line stacking. Each column shows all the transitions of a particular species used in the line stacking are shown as grey-filled histograms. The black histograms represent the resulting stacked spectrum of each molecule for comparison.
}
\label{fig:stack_all}
\end{figure*}

To unveil the complexity of the line profiles, we stack spectra from manually selected transitions using SciPy and NumPy; these transitions are selected with similar line profiles and not contaminated by other lines.
Figure \ref{fig:stack_all} shows the individual line profiles in comparison to the stacked line profiles presented in Fig. \ref{fig:stack}.
These transitions are selected as they share similar normalized spectrum profiles; we compare the averaged profiles with the candidate transition to see if they look similar. If yes, the transition is added for stacking.
In the off-peak velocities, the continuum-subtracted fluxes sometimes do not well reach zero since they can still be affected by contamination from other molecular line transitions.

\section{Line fitting}
\label{app:fitting}
\begin{figure*}
\centering
\includegraphics[width=1\textwidth]{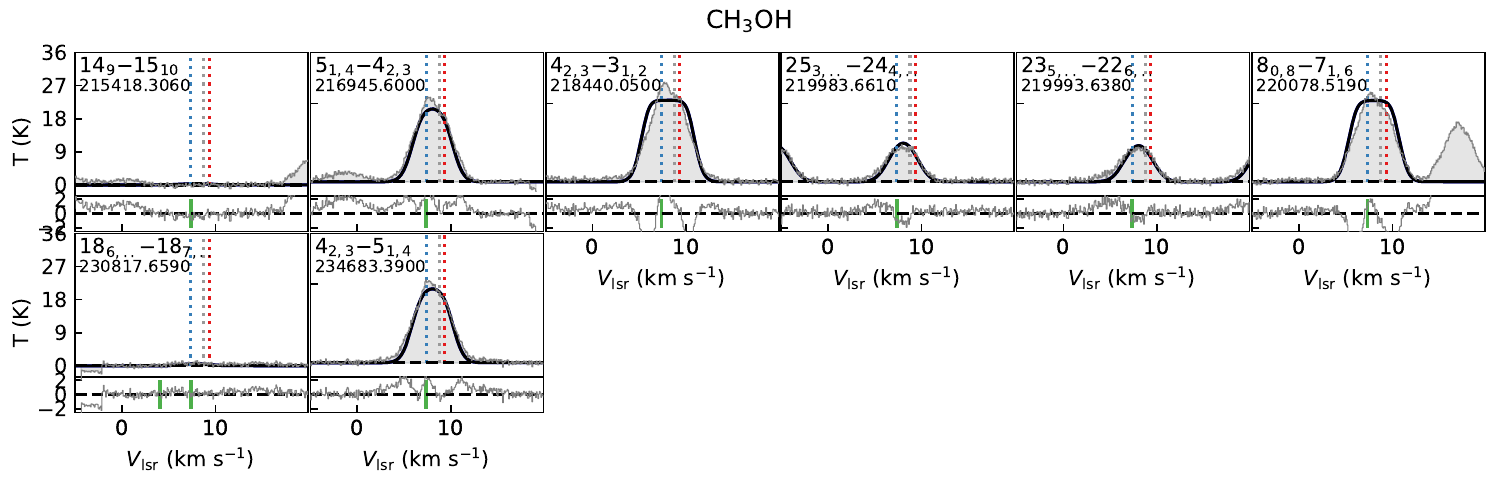}
\caption{Modeled \ce{CH3OH} line profiles overlaid on the observed spectra. The three vertical dashed lines show the central velocities of VLA4A (blue), VLA4B (red), and the circumbinary system (grey). The sub-panel at the bottom shows the residual from the best-fit. The green bars mark the velocities of VLA4A (7.36 km s$^{-1}$) for all transitions used in the model of \ce{CH3OH}.
}
\label{fig:Fitting_ch3oh}
\end{figure*}

\begin{figure*}
\centering
\includegraphics[width=1\textwidth]{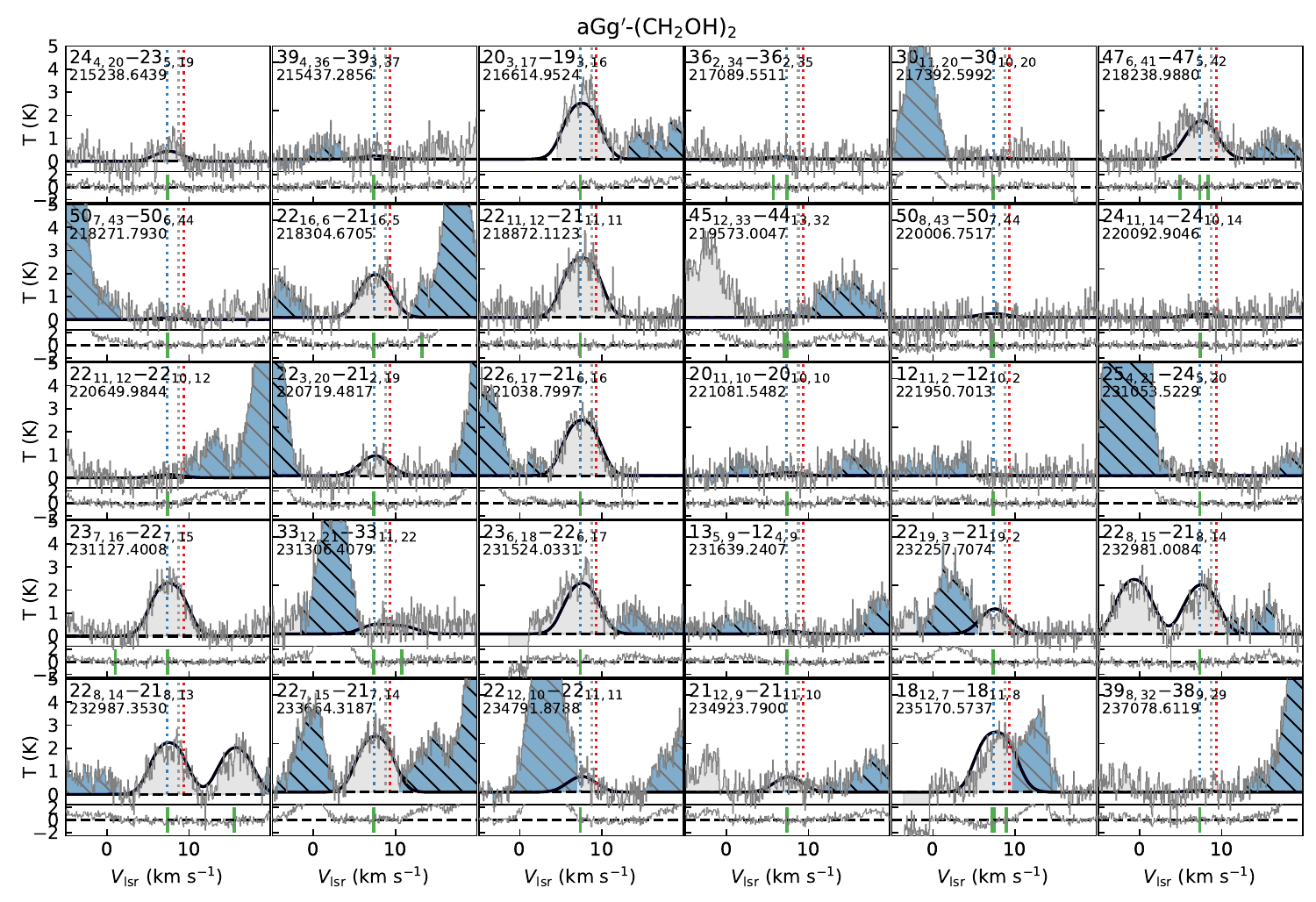}
\caption{Same as Figure \ref{fig:Fitting_ch3oh} but for aGg'-\ce{(CH2OH)2}. The hatch indicates the regions that are masked for the fitting.
}
\label{fig:Fitting_ch2oh2}
\end{figure*}

\begin{figure*}
\centering
\includegraphics[width=1\textwidth]{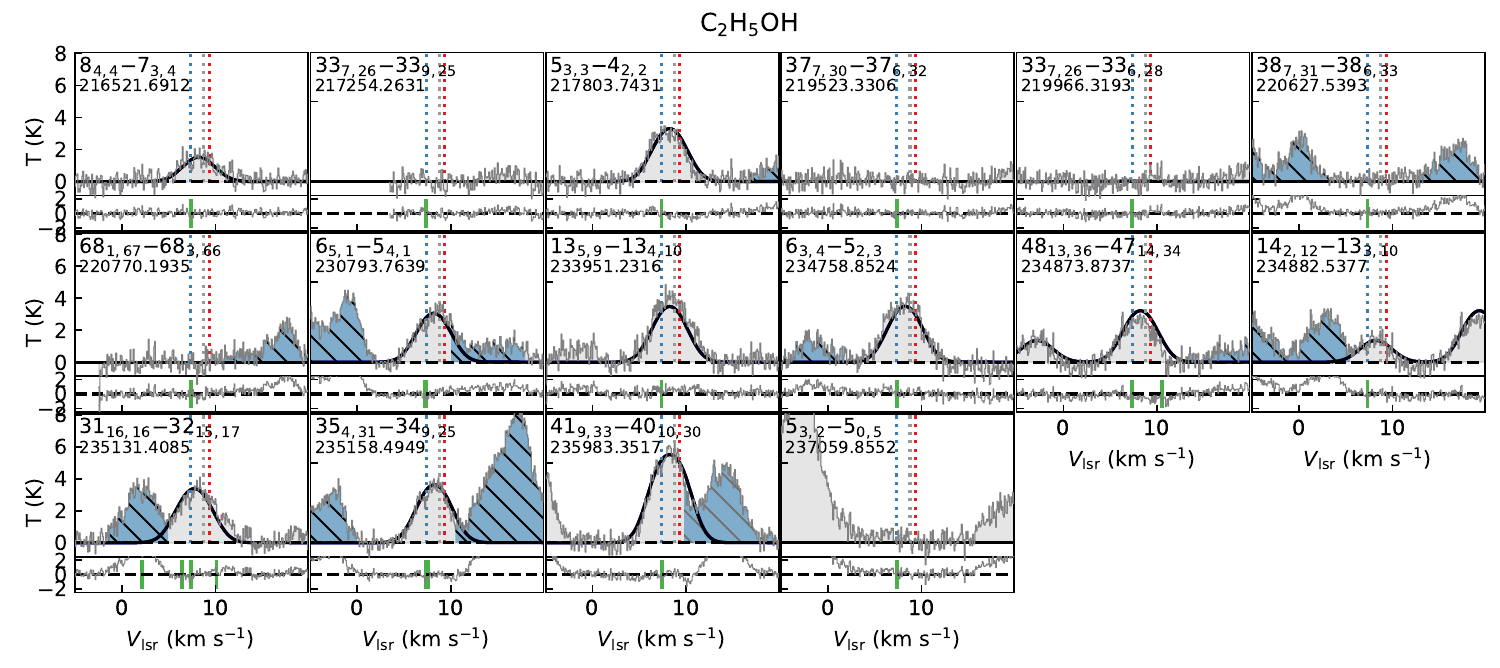}
\caption{Same as Figure \ref{fig:Fitting_ch2oh2} but for \ce{C2H5OH}.
}
\label{fig:Fitting_c2h5oh}
\end{figure*}

\begin{figure*}
\centering
\includegraphics[width=1\textwidth]{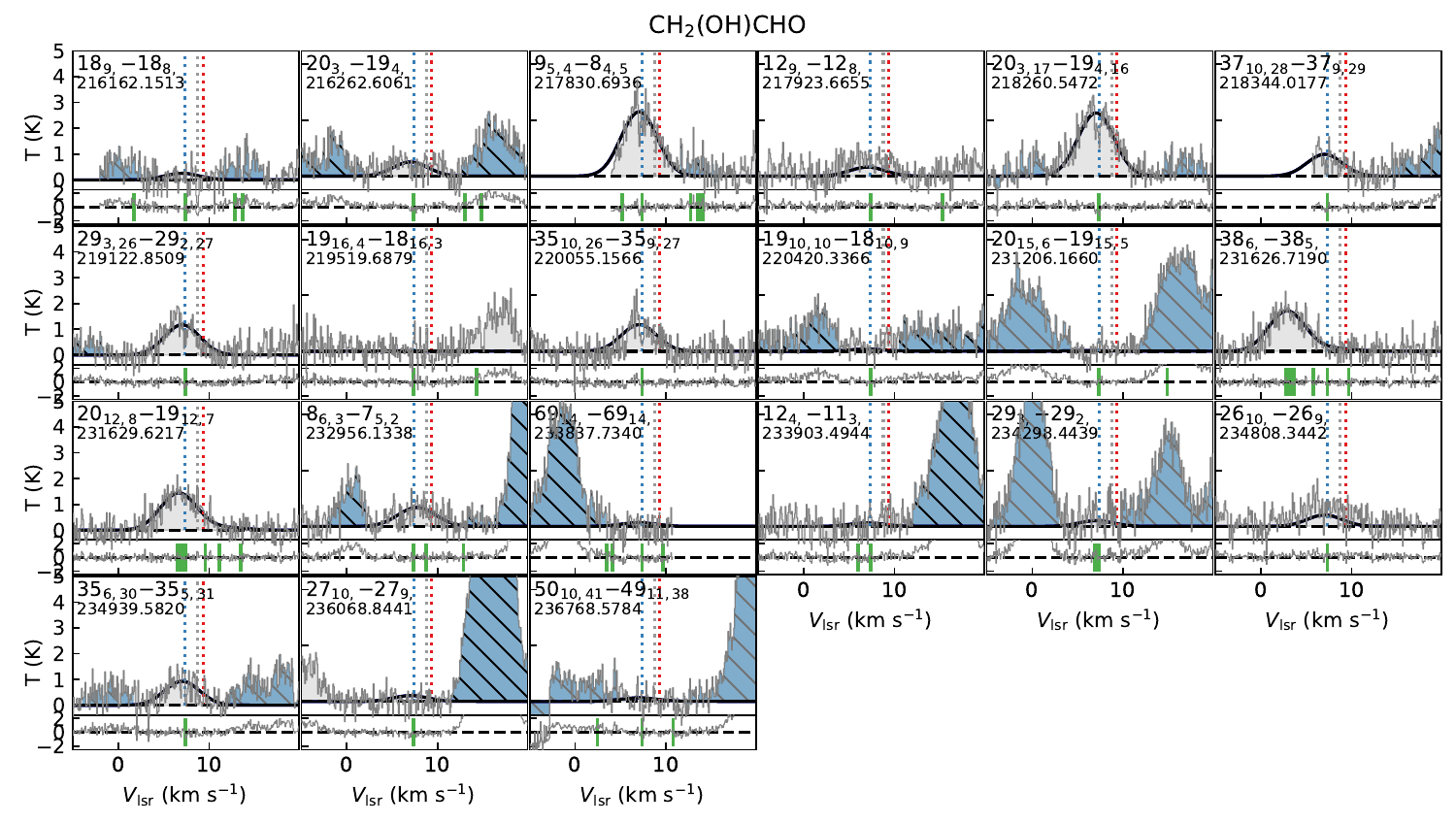}
\caption{Same as Figure \ref{fig:Fitting_ch2oh2} but for \ce{CH2(OH)CHO}.
}
\label{fig:Fitting_ch2ohcho}
\end{figure*}

\begin{figure*}
\centering
\includegraphics[width=1\textwidth]{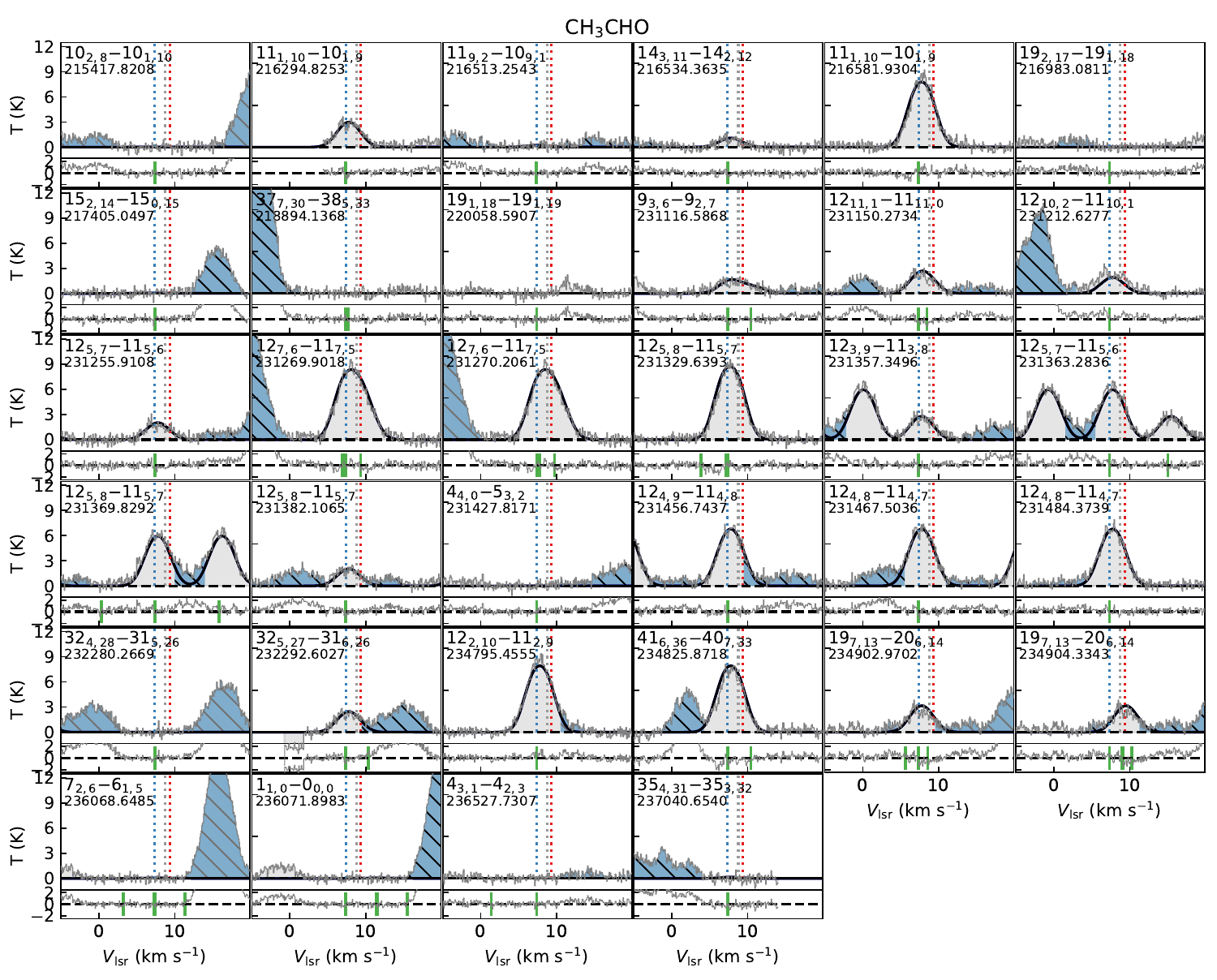}
\caption{Same as Figure \ref{fig:Fitting_ch2oh2} but for \ce{CH3CHO}.
}
\label{fig:Fitting_ch3cho}
\end{figure*}

\begin{figure*}
\centering
\includegraphics[width=1\textwidth]{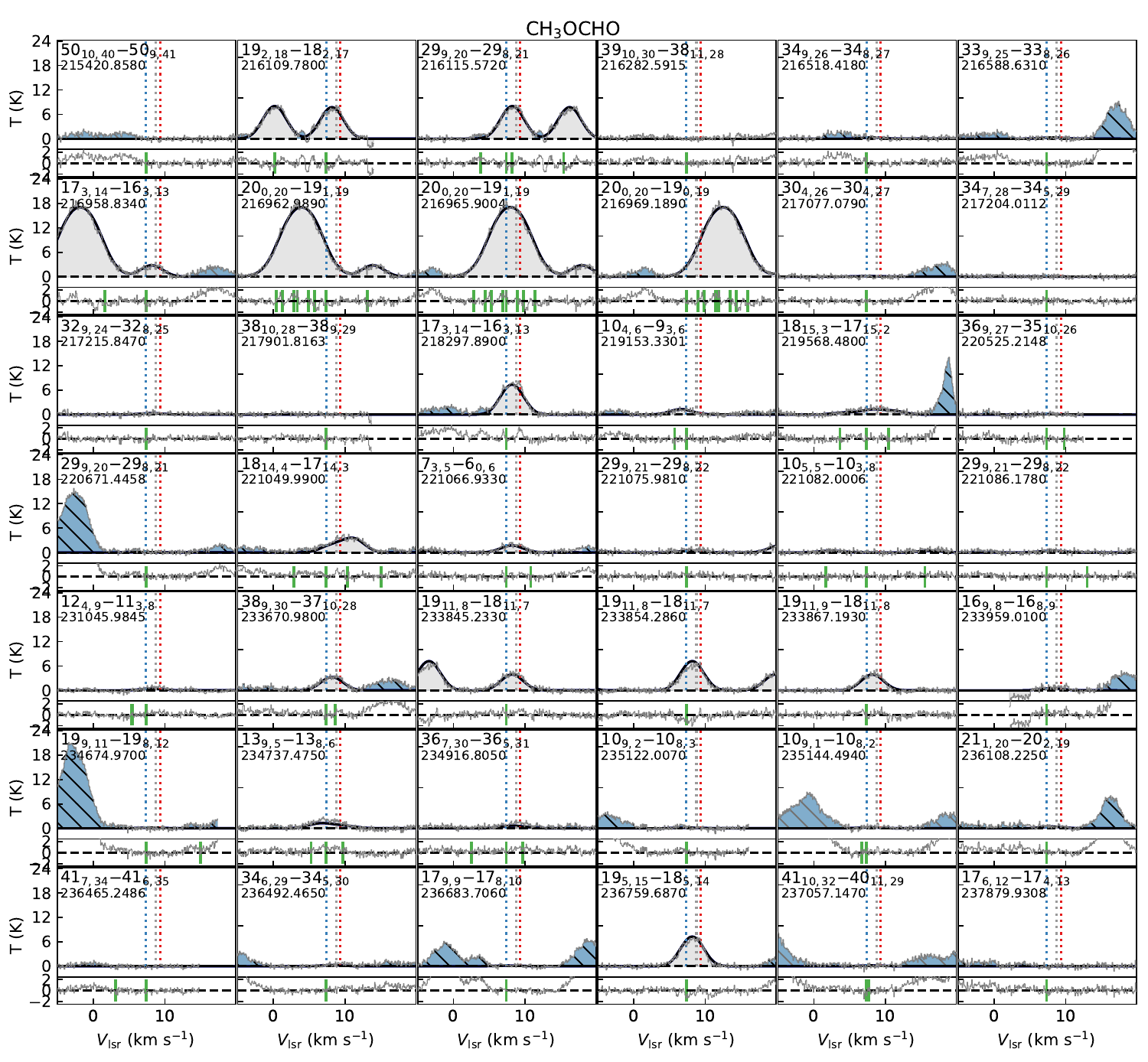}
\caption{Same as Figure \ref{fig:Fitting_ch2oh2} but for \ce{CH3OCHO}.
}
\label{fig:Fitting_ch3ocho}
\end{figure*}

\begin{figure*}
\centering
\includegraphics[width=1\textwidth]{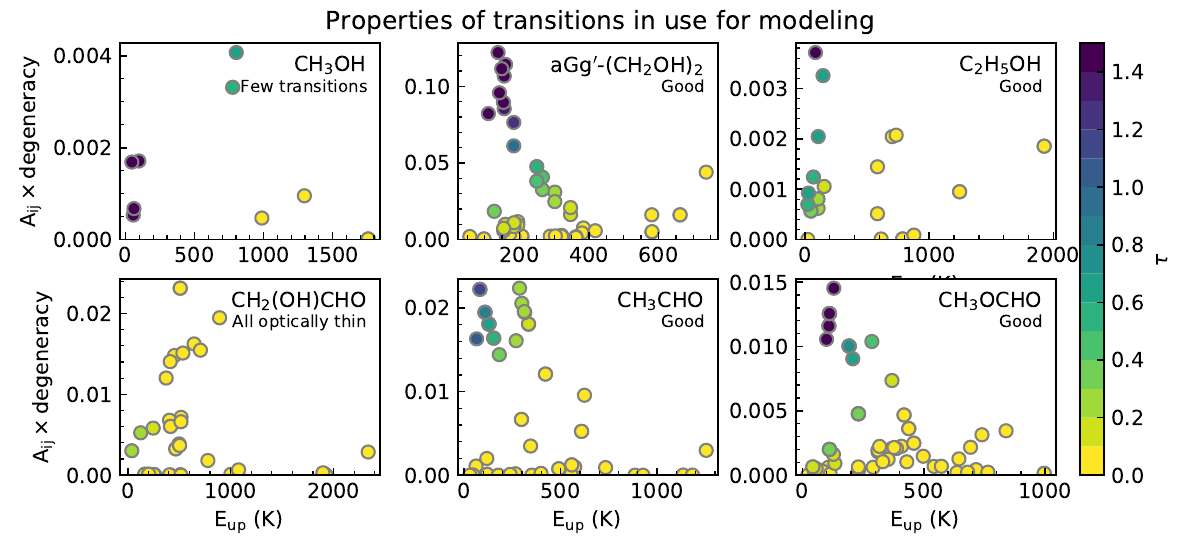}
\caption{Properties of the transitions in use for the modeling. The color of data points indicates the optical depth for each transition from the single velocity component model. The label ``Good'' indicate that the transitions in use well sample the energy levels to derive the temperature and cover optically thick emission to derive the source size.
}
\label{fig:trans}
\end{figure*}

\Cref{fig:Fitting_ch3oh,fig:Fitting_ch2oh2,fig:Fitting_c2h5oh,fig:Fitting_ch2ohcho,fig:Fitting_ch3cho,fig:Fitting_ch3ocho} show the best-fit profiles of the one-component fitting for all transitions in use (see section \ref{sec:oneGau}).
Figure \ref{fig:trans} shows the relative intensity as a function of the upper energy level for those transitions involved in the fitting; more importantly, the colors represent the optical depths from the single-component fitting.
Optically thick lines are required in order to derive the source size (see section \ref{sec:oneGau}). 
On the other hand, a broad range of $E_{\rm up}$ will constrain the physical conditions and perhaps identify different temperature components. Also, a large number of transitions in use is crucial, especially for multi-component fitting.

For each COM, we start by fitting one component and add an extra component with each iteration of the fit. For each iteration, $\chi^2$ and $\Delta$AIC are calculated to help evaluate if adding one more component will improve the fit (Table \ref{tab:aic}). We stop adding additional components when the $\chi^2$ does not significantly improve comparing with the previous iteration and/or the fitting found an unreasonable component. Take \ce{C2H5OH} as an example. 
We stop adding a new component at the 2-component model because
the 3rd component contributes emission at about the noise level with an extreme high optical depth ($9\times10^{22}$ cm$^{-2}$). 
Whether this component is real or not is difficult to judge with the current data set and severe line contamination in this source. Therefore, we adopt the 2-component model for \ce{C2H5OH}. 
After the numbers of components are decided we conduct MCMC fitting using {\tt emcee} as well as the one component fitting.

\begin{table*}
    \centering
    \caption{Transitions for the fitting - \ce{CH3OH}}
    \begin{tabular}{cccccccccccc}
    \hline\hline
Rest frequency
& Transition
& E$_{\rm up}$
& A$_{\rm ij}$
& upper state
& $\tau_{\rm 1 comp.}$
& Gau. fitting/stacked
& uv-space center
& S/N
\\
(MHz)
&	
& (K)
& (s$^{-1}$)
& degeneracy
& 
& 
&
& \\
\hline
215418.3060 & 14$_{9}$$-$15$_{10}$  & 1295.6    & 3.29e-05  & 29    & 0.018 & - & - & 2.2   \\
216945.6000 & 5$_{1,4}$$-$4$_{2,3}$ & 55.9  & 1.21e-05  & 44    & 2.247 & yes   & yes   & 58.5  \\
218440.0500 & 4$_{2,3}$$-$3$_{1,2}$ & 45.5  & 4.69e-05  & 36    & 7.334 & yes   & yes   & 64.8  \\
219983.6610 & 25$_{3,..}$$-$24$_{4,..}$ & 802.2 & 2.04e-05  & 200   & 0.641 & yes   & yes   & 24.0  \\
219993.6380 & 23$_{5,..}$$-$22$_{6,..}$ & 775.9 & 1.75e-05  & 190   & 0.585 & yes   & yes   & 22.4  \\
220078.5190 & 8$_{0,8}$$-$7$_{1,6}$ & 96.6  & 2.52e-05  & 68    & 5.862 & yes   & yes   & 57.8  \\
230817.6590 & 18$_{6,..}$$-$18$_{7,..}$ & 987.6 & 3.13e-06  & 150   & 0.030 & - & - & 4.5   \\
230820.1620 & 30$_{9,}$$-$31$_{11,}$    & 1757.7    & 5.29e-08  & 240   & 0.000 & - & - & blend \\
230820.1620 & 30$_{9,}$$-$31$_{11,}$    & 1757.7    & 5.29e-08  & 240   & 0.000 & - & - & blend \\
234683.3900 & 4$_{2,3}$$-$5$_{1,4}$ & 60.9  & 1.87e-05  & 36    & 2.379 & yes & - & 57.8  \\
\hline
    \end{tabular}
\tablefoot{The transitions used for LTE model fitting.
Column 1-5: The transition information obtained from CDMS using the XCLASS task {\tt ListDatabase}.
Column 6: Optical depths derived from the one-component LTE model fitting, which are also shown in Figure \ref{fig:trans}.
Column 7: The transitions that are used in the Gaussian fitting (Figure \ref{fig:gau}) and the stacked process (Figure \ref{fig:stack_all}).
Column 8: The transitions used for 2D Gaussian fitting in the uv-space for the emitting centers in Figure \ref{fig:uv_cen}.
Column 9: The signal to noise ratio obtained given the peak pixel value at the resolution of $\sim0.08$ km s$^{-1}$. ``blend'' indicates that the transition is blended with nearby transitions of the same molecule, for which we only list one S/N for the group (see corresponding Figure \ref{fig:Fitting_ch3oh}).
}
    \label{tab:tran_ch3oh}
\end{table*}

\begin{table*}
    \centering
    \caption{Transitions for the fitting - aGg'-\ce{(CH2OH)2}}
    \begin{tabular}{cccccccccccc}
    \hline\hline
Rest frequency
& Transition
& E$_{\rm up}$
& A$_{\rm ij}$
& upper state
& $\tau_{\rm 1 comp.}$
& Gau. fitting/stacked
& uv-space center
& S/N
\\
(MHz)
&	
& (K)
& (s$^{-1}$)
& degeneracy
& 
& 
&
& \\
\hline
215238.6439 & 24$_{4,20}$$-$23$_{5,19}$ & 159.4 & 2.26e-05  & 440   & 0.184 & - & - & 3.6   \\
215437.2856 & 39$_{4,36}$$-$39$_{3,37}$ & 384.0 & 1.38e-05  & 550   & 0.055 & - & - & 2.6   \\
216614.9524 & 20$_{3,17}$$-$19$_{3,16}$ & 110.8 & 2.22e-04  & 370   & 1.835 & yes   & yes   & 9.5   \\
217089.5511 & 36$_{2,34}$$-$36$_{2,35}$ & 321.3 & 4.14e-06  & 660   & 0.025 & - & - & 3.0   \\
217090.7383 & 36$_{3,34}$$-$36$_{1,35}$ & 321.3 & 4.15e-06  & 510   & 0.020 & - & - & blend \\
217392.5992 & 30$_{11,20}$$-$30$_{10,20}$   & 288.6 & 5.20e-06  & 430   & 0.024 & - & - & 1.5   \\
218238.2595 & 47$_{6,41}$$-$47$_{5,42}$ & 580.6 & 2.45e-05  & 660   & 0.051 & - & - & blend \\
218238.9880 & 22$_{17,6}$$-$21$_{17,5}$ & 266.2 & 1.02e-04  & 400   & 0.470 & yes   & yes   & 6.7   \\
218238.9880 & 22$_{17,5}$$-$21$_{17,4}$ & 266.2 & 1.02e-04  & 320   & 0.376 & yes   & yes   & 6.7   \\
218240.7716 & 33$_{7,26}$$-$32$_{8,24}$ & 303.4 & 3.97e-06  & 600   & 0.024 & - & - & blend \\
218271.7930 & 50$_{7,43}$$-$50$_{6,44}$ & 661.7 & 2.27e-05  & 710   & 0.036 & - & - & 2.5   \\
218304.6705 & 22$_{16,6}$$-$21$_{16,5}$ & 250.0 & 1.19e-04  & 320   & 0.471 & - & - & 6.8   \\
218304.6705 & 22$_{16,7}$$-$21$_{16,6}$ & 250.0 & 1.19e-04  & 400   & 0.588 & - & - & 6.8   \\
218872.1123 & 22$_{11,12}$$-$21$_{11,11}$   & 183.8 & 1.91e-04  & 400   & 1.235 & yes   & yes   & 8.8   \\
218872.1125 & 22$_{11,11}$$-$21$_{11,10}$   & 183.8 & 1.91e-04  & 320   & 0.988 & - & - & blend \\
219573.0047 & 45$_{12,33}$$-$44$_{13,32}$   & 581.1 & 5.98e-06  & 820   & 0.015 & - & - & 2.1   \\
219573.2390 & 45$_{12,33}$$-$44$_{13,32}$   & 581.4 & 8.05e-06  & 640   & 0.016 & - & - & blend \\
220006.7517 & 50$_{8,43}$$-$50$_{7,44}$ & 662.1 & 2.28e-05  & 710   & 0.035 & - & - & 2.2   \\
220006.9692 & 37$_{8,30}$$-$36$_{9,27}$ & 379.8 & 6.35e-06  & 680   & 0.031 & - & - & blend \\
220092.8238 & 24$_{11,14}$$-$24$_{10,14}$   & 207.2 & 5.15e-06  & 340   & 0.025 & - & - & blend \\
220092.9046 & 24$_{11,13}$$-$24$_{10,15}$   & 207.2 & 5.15e-06  & 440   & 0.033 & - & - & 2.4   \\
220649.9710 & 22$_{11,12}$$-$22$_{10,12}$   & 184.1 & 5.01e-06  & 320   & 0.025 & - & - & blend \\
220649.9844 & 22$_{11,11}$$-$22$_{10,13}$   & 184.1 & 5.01e-06  & 400   & 0.032 & - & - & 1.9   \\
220719.4817 & 22$_{3,20}$$-$21$_{2,19}$ & 127.8 & 5.77e-05  & 320   & 0.370 & - & - & 3.7   \\
221038.7997 & 22$_{6,17}$$-$21$_{6,16}$ & 142.6 & 2.39e-04  & 400   & 1.800 & yes   & yes   & 8.5   \\
221081.5463 & 20$_{11,10}$$-$20$_{10,10}$   & 163.0 & 4.79e-06  & 290   & 0.024 & - & - & blend \\
221081.5482 & 20$_{11,9}$$-$20$_{10,11}$    & 163.0 & 4.79e-06  & 370   & 0.031 & - & - & 2.6   \\
221950.7013 & 12$_{11,2}$$-$12$_{10,2}$ & 98.1  & 1.98e-06  & 180   & 0.008 & - & - & 2.4   \\
221950.7013 & 12$_{11,1}$$-$12$_{10,3}$ & 98.1  & 1.98e-06  & 220   & 0.010 & - & - & 2.4   \\
231053.5229 & 25$_{4,21}$$-$24$_{5,20}$ & 171.7 & 7.64e-06  & 460   & 0.054 & - & - & 1.9   \\
231127.4008 & 23$_{7,16}$$-$22$_{7,15}$ & 160.2 & 2.72e-04  & 420   & 1.829 & - & - & 8.6   \\
231303.7540 & 33$_{12,21}$$-$33$_{11,22}$   & 347.0 & 3.47e-05  & 470   & 0.120 & - & - & blend \\
231306.4079 & 33$_{12,22}$$-$33$_{11,23}$   & 347.0 & 3.48e-05  & 600   & 0.153 & - & - & 3.8   \\
231524.0331 & 23$_{6,18}$$-$22$_{6,17}$ & 154.1 & 2.71e-04  & 330   & 1.465 & yes   & yes   & 9.1   \\
231639.2407 & 13$_{5,9}$$-$12$_{4,9}$   & 57.4  & 8.34e-06  & 240   & 0.049 & - & - & 2.5   \\
232257.7074 & 22$_{19,3}$$-$21$_{19,2}$ & 301.8 & 7.74e-05  & 400   & 0.273 & - & - & 4.7   \\
232257.7074 & 22$_{19,4}$$-$21$_{19,3}$ & 301.8 & 7.74e-05  & 320   & 0.218 & - & - & 4.7   \\
232981.0084 & 22$_{8,15}$$-$21$_{8,14}$ & 156.3 & 2.67e-04  & 320   & 1.367 & yes   & yes   & 7.4   \\
232987.3530 & 22$_{8,14}$$-$21$_{8,13}$ & 156.3 & 2.67e-04  & 400   & 1.708 & yes   & yes   & 7.7   \\
233664.3187 & 22$_{7,15}$$-$21$_{7,14}$ & 149.1 & 2.79e-04  & 400   & 1.828 & - & - & 8.5   \\
234791.8788 & 22$_{12,10}$$-$22$_{11,11}$   & 195.4 & 2.92e-05  & 400   & 0.157 & - & - & 6.2   \\
234791.8791 & 22$_{12,11}$$-$22$_{11,12}$   & 195.4 & 2.92e-05  & 320   & 0.125 & - & - & blend \\
234923.7899 & 21$_{12,9}$$-$21$_{11,10}$    & 184.6 & 2.82e-05  & 300   & 0.118 & - & - & blend \\
234923.7900 & 21$_{12,10}$$-$21$_{11,11}$   & 184.6 & 2.82e-05  & 390   & 0.154 & - & - & 3.6   \\
235169.3027 & 18$_{12,7}$$-$18$_{11,8}$ & 154.7 & 2.23e-05  & 330   & 0.116 & - & - & blend \\
235169.3027 & 18$_{12,6}$$-$18$_{11,7}$ & 154.7 & 2.23e-05  & 260   & 0.092 & - & - & blend \\
235170.4758 & 51$_{13,39}$$-$51$_{12,40}$   & 737.1 & 4.73e-05  & 930   & 0.062 & - & - & blend \\
235170.5737 & 23$_{3,21}$$-$22$_{3,20}$ & 138.7 & 2.91e-04  & 420   & 2.063 & - & - & 8.0   \\
235170.6954 & 39$_{1,38}$$-$39$_{1,39}$ & 362.6 & 2.56e-06  & 710   & 0.012 & - & - & blend \\
235170.7008 & 39$_{2,38}$$-$39$_{0,39}$ & 362.6 & 2.56e-06  & 550   & 0.009 & - & - & blend \\
237078.6119 & 39$_{8,32}$$-$38$_{9,29}$ & 418.2 & 8.11e-06  & 710   & 0.030 & - & - & 2.3   \\
\hline
    \end{tabular}
\tablefoot{Same for Table \ref{tab:tran_ch3oh} but for aGg'-\ce{(CH2OH)2} with the corresponding figure as Figure \ref{fig:Fitting_ch2oh2}.
}
    \label{tab:tran_ch2oh2}
\end{table*}

\begin{table*}
    \centering
    \caption{Transitions for the fitting - \ce{C2H5OH}}
    \begin{tabular}{cccccccccccc}
    \hline\hline
Rest frequency
& Transition
& E$_{\rm up}$
& A$_{\rm ij}$
& upper state
& $\tau_{\rm 1 comp.}$
& Gau. fitting/stacked
& uv-space center
& S/N
\\
(MHz)
&	
& (K)
& (s$^{-1}$)
& degeneracy
& 
& 
&
&\\
\hline
216521.6912 & 8$_{4,4}$$-$7$_{3,4}$ & 106.3 & 3.65e-05  & 17    & 0.225 & - & - & 5.9   \\
217254.2631 & 33$_{7,26}$$-$33$_{9,25}$ & 583.9 & 7.66e-06  & 67    & 0.001 & - & - & 2.3   \\
217803.7431 & 5$_{3,3}$$-$4$_{2,2}$ & 23.9  & 6.34e-05  & 11    & 0.582 & yes   & yes   & 9.2   \\
219523.3306 & 37$_{7,30}$$-$37$_{6,32}$ & 702.7 & 2.72e-05  & 75    & 0.002 & - & - & 1.4   \\
219966.3193 & 33$_{7,26}$$-$33$_{6,28}$ & 583.9 & 2.16e-05  & 67    & 0.004 & - & - & 2.7   \\
220627.5393 & 38$_{7,31}$$-$38$_{6,33}$ & 734.5 & 2.69e-05  & 77    & 0.001 & - & - & 2.2   \\
220770.1935 & 68$_{1,67}$$-$68$_{3,66}$ & 1925.2    & 1.32e-05  & 140   & 0.000 & - & - & 2.8   \\
220770.2093 & 68$_{2,67}$$-$68$_{2,66}$ & 1925.2    & 1.32e-05  & 140   & 0.000 & - & - & blend \\
230793.7639 & 6$_{5,1}$$-$5$_{4,1}$ & 104.8 & 6.20e-05  & 13    & 0.263 & yes   & - & 11.6  \\
230793.8643 & 6$_{5,2}$$-$5$_{4,2}$ & 104.8 & 6.20e-05  & 13    & 0.263 & yes   & - & blend \\
233951.2316 & 13$_{5,9}$$-$13$_{4,10}$  & 107.9 & 7.57e-05  & 27    & 0.628 & yes   & yes   & 12.2  \\
234758.8524 & 6$_{3,4}$$-$5$_{2,3}$ & 28.9  & 7.09e-05  & 13    & 0.632 & - & - & 11.8  \\
234871.3761 & 48$_{13,36}$$-$47$_{14,34}$   & 1245.2    & 9.78e-06  & 97    & 0.000 & - & - & blend \\
234871.3864 & 48$_{13,35}$$-$47$_{14,33}$   & 1245.2    & 9.78e-06  & 97    & 0.000 & - & - & blend \\
234873.8737 & 9$_{5,5}$$-$9$_{4,6}$ & 69.2  & 6.54e-05  & 19    & 0.563 & - & - & 8.5   \\
234882.5377 & 14$_{2,12}$$-$13$_{3,10}$ & 155.2 & 3.64e-05  & 29    & 0.198 & - & - & 5.6   \\
235129.2333 & 31$_{16,16}$$-$32$_{15,17}$   & 786.9 & 1.89e-07  & 63    & 0.000 & - & - & blend \\
235129.2333 & 31$_{16,15}$$-$32$_{15,18}$   & 786.9 & 1.89e-07  & 63    & 0.000 & - & - & blend \\
235131.4085 & 6$_{5,1}$$-$6$_{4,2}$ & 49.0  & 4.36e-05  & 13    & 0.315 & - & - & 11.2  \\
235132.1549 & 6$_{5,2}$$-$6$_{4,3}$ & 49.0  & 4.36e-05  & 13    & 0.315 & - & - & blend \\
235158.3101 & 35$_{4,31}$$-$34$_{9,25}$ & 613.4 & 1.74e-07  & 71    & 0.000 & - & - & blend \\
235158.4949 & 14$_{0,14}$$-$13$_{1,12}$ & 146.4 & 1.12e-04  & 29    & 0.667 & - & - & 11.1  \\
235983.3336 & 41$_{9,33}$$-$40$_{10,30}$    & 877.2 & 1.07e-06  & 83    & 0.000 & - & - & blend \\
235983.3517 & 14$_{1,14}$$-$13$_{0,13}$ & 85.6  & 1.28e-04  & 29    & 1.409 & yes   & yes   & 17.5  \\
237059.8552 & 5$_{3,2}$$-$5$_{0,5}$ & 23.9  & 2.25e-07  & 11    & 0.002 & - & - & 2.5   \\
\hline
    \end{tabular}
\tablefoot{Same for Table \ref{tab:tran_ch3oh} but for \ce{C2H5OH} with the corresponding figure as Figure \ref{fig:Fitting_c2h5oh}.
}
    \label{tab:tran_c2h5oh}
\end{table*}

\begin{table*}
    \centering
    \caption{Transitions for the fitting - \ce{CH2(OH)CHO}}
    \begin{tabular}{cccccccccccc}
    \hline\hline
Rest frequency
& Transition
& E$_{\rm up}$
& A$_{\rm ij}$
& upper state
& $\tau_{\rm 1 comp.}$
& Gau. fitting/stacked
& uv-space center
& S/N
\\
(MHz)
&	
& (K)
& (s$^{-1}$)
& degeneracy
& 
&
& 
&\\
\hline
216162.1513 & 18$_{9,}$$-$18$_{8,}$ & 518.6 & 1.93e-04  & 37    & 0.023 & - & - & 3.0   \\
216262.6061 & 20$_{3,}$$-$19$_{4,}$ & 406.4 & 1.66e-04  & 41    & 0.045 & - & - & 2.9   \\
217830.6936 & 9$_{5,4}$$-$8$_{4,5}$ & 40.2  & 1.59e-04  & 19    & 0.220 & yes   & yes   & 9.1   \\
217832.2544 & 67$_{18,}$$-$68$_{15,}$   & 1909.6    & 1.48e-06  & 140   & 0.000 & - & - & blend \\
217923.6615 & 12$_{9,}$$-$12$_{8,}$ & 467.2 & 1.31e-04  & 25    & 0.014 & - & - & blend \\
217923.6655 & 12$_{9,}$$-$12$_{8,}$ & 467.2 & 1.31e-04  & 25    & 0.014 & - & - & 3.3   \\
218260.5472 & 20$_{3,17}$$-$19$_{4,16}$ & 126.1 & 1.28e-04  & 41    & 0.216 & yes   & yes   & 7.8   \\
218344.0177 & 37$_{10,28}$$-$37$_{9,29}$    & 454.9 & 1.97e-04  & 75    & 0.070 & yes   & yes   & 3.8   \\
219122.8509 & 29$_{3,26}$$-$29$_{2,27}$ & 247.8 & 9.87e-05  & 59    & 0.107 & yes   & yes   & 5.0   \\
219519.6879 & 19$_{16,4}$$-$18$_{16,3}$ & 260.9 & 1.20e-06  & 39    & 0.001 & - & - & 2.7   \\
219519.6879 & 19$_{16,3}$$-$18$_{16,2}$ & 260.9 & 1.20e-06  & 39    & 0.001 & - & - & 2.7   \\
220055.1566 & 35$_{10,26}$$-$35$_{9,27}$    & 413.5 & 1.98e-04  & 71    & 0.086 & yes   & yes   & 5.6   \\
220420.3366 & 19$_{10,10}$$-$18$_{10,9}$    & 166.3 & 3.02e-06  & 39    & 0.004 & - & - & 2.2   \\
220420.3451 & 19$_{10,9}$$-$18$_{10,8}$ & 166.3 & 3.02e-06  & 39    & 0.004 & - & - & blend \\
231206.1660 & 20$_{15,6}$$-$19$_{15,5}$ & 253.2 & 2.11e-06  & 41    & 0.001 & - & - & 2.5   \\
231206.1660 & 20$_{15,5}$$-$19$_{15,4}$ & 253.2 & 2.11e-06  & 41    & 0.001 & - & - & 2.5   \\
231624.9077 & 38$_{6,}$$-$38$_{5,}$ & 892.9 & 2.53e-04  & 77    & 0.005 & - & - & blend \\
231626.7190 & 20$_{12,8}$$-$19$_{12,7}$ & 204.0 & 3.10e-06  & 41    & 0.003 & - & - & 5.1   \\
231626.7190 & 20$_{12,9}$$-$19$_{12,8}$ & 204.0 & 3.10e-06  & 41    & 0.003 & - & - & 5.1   \\
231627.9025 & 21$_{2,}$$-$20$_{2,}$ & 411.6 & 9.83e-07  & 43    & 0.000 & - & - & blend \\
231629.5555 & 44$_{11,}$$-$45$_{8,}$    & 1001.6    & 1.15e-06  & 89    & 0.000 & - & - & blend \\
231629.6217 & 23$_{0,}$$-$22$_{1,}$ & 510.1 & 4.92e-04  & 47    & 0.068 & - & - & 6.3   \\
231629.8905 & 23$_{1,}$$-$22$_{1,}$ & 510.1 & 1.01e-06  & 47    & 0.000 & - & - & blend \\
231630.0690 & 23$_{0,}$$-$22$_{0,}$ & 510.1 & 1.01e-06  & 47    & 0.000 & - & - & blend \\
231630.2222 & 23$_{1,}$$-$22$_{0,}$ & 510.1 & 4.92e-04  & 47    & 0.068 & yes   & yes   & blend \\
232955.0612 & 8$_{6,3}$$-$7$_{5,2}$ & 416.3 & 3.55e-04  & 17    & 0.032 & - & - & blend \\
232956.1338 & 8$_{6,2}$$-$7$_{5,3}$ & 416.3 & 3.55e-04  & 17    & 0.032 & - & - & 3.7   \\
233835.9409 & 69$_{14,}$$-$69$_{14,}$   & 1928.6    & 3.59e-08  & 140   & 0.000 & - & - & blend \\
233837.7340 & 39$_{10,29}$$-$38$_{11,28}$   & 499.3 & 4.90e-05  & 79    & 0.012 & - & - & 1.4   \\
233840.2668 & 62$_{38,}$$-$63$_{37,}$   & 2335.9    & 2.39e-05  & 120   & 0.000 & - & - & blend \\
233840.7525 & 31$_{24,}$$-$32$_{23,}$   & 1077.3    & 1.05e-05  & 63    & 0.000 & - & - & blend \\
233903.4944 & 12$_{4,}$$-$11$_{3,}$ & 504.7 & 1.47e-04  & 25    & 0.011 & - & - & 2.2   \\
233904.5100 & 69$_{17,}$$-$70$_{14,}$   & 1896.0    & 2.10e-06  & 140   & 0.000 & - & - & blend \\
233904.5390 & 20$_{3,}$$-$19$_{3,}$ & 406.4 & 1.00e-06  & 41    & 0.000 & - & - & blend \\
234298.4439 & 29$_{3,}$$-$29$_{2,}$ & 517.5 & 1.12e-04  & 59    & 0.018 & - & - & 3.2   \\
234298.7988 & 29$_{3,}$$-$29$_{1,}$ & 517.5 & 2.17e-08  & 59    & 0.000 & - & - & blend \\
234808.3442 & 26$_{10,}$$-$26$_{9,}$    & 536.3 & 2.85e-04  & 53    & 0.036 & - & - & 4.7   \\
234939.5820 & 35$_{6,30}$$-$35$_{5,31}$ & 373.6 & 1.69e-04  & 71    & 0.084 & - & - & 4.4   \\
236068.8441 & 27$_{10,}$$-$27$_{9,}$    & 645.0 & 2.95e-04  & 55    & 0.019 & - & - & 1.9   \\
236765.8791 & 50$_{10,41}$$-$49$_{11,38}$   & 778.9 & 1.82e-05  & 100   & 0.001 & - & - & blend \\
236768.5784 & 26$_{10,}$$-$26$_{9,}$    & 706.4 & 2.92e-04  & 53    & 0.012 & - & - & 2.6   \\
\hline
    \end{tabular}
\tablefoot{Same for Table \ref{tab:tran_ch3oh} but for \ce{CH2(OH)CHO} with the corresponding figure as Figure \ref{fig:Fitting_ch2ohcho}.
}
    \label{tab:tran_ch2ohcho}
\end{table*}

\begin{table*}
    \centering
    \caption{Transitions for the fitting - \ce{CH3CHO}}
    \begin{tabular}{cccccccccccc}
    \hline\hline
Rest frequency
& Transition
& E$_{\rm up}$
& A$_{\rm ij}$
& upper state
& $\tau_{\rm 1 comp.}$
& Gau. fitting/stacked
& uv-space center
& S/N
\\
(MHz)
&	
& (K)
& (s$^{-1}$)
& degeneracy
& 
&
& 
& \\
\hline
215417.8208 & 10$_{2,8}$$-$10$_{1,10}$  & 266.3 & 4.05e-06  & 42    & 0.003 & - & - & 2.2   \\
216294.8253 & 11$_{1,10}$$-$10$_{1,9}$  & 269.6 & 3.50e-04  & 46    & 0.283 & - & - & blend \\
216513.2543 & 11$_{9,2}$$-$10$_{9,1}$   & 608.7 & 1.14e-04  & 46    & 0.010 & - & - & 2.3   \\
216513.2579 & 11$_{9,3}$$-$10$_{9,2}$   & 608.7 & 1.14e-04  & 46    & 0.010 & - & - & blend \\
216534.3635 & 14$_{3,11}$$-$14$_{2,12}$ & 117.7 & 3.45e-05  & 58    & 0.093 & yes   & yes   & 4.6   \\
216581.9304 & 11$_{1,10}$$-$10$_{1,9}$  & 64.9  & 3.55e-04  & 46    & 1.064 & - & - & blend \\
216983.0811 & 19$_{2,17}$$-$19$_{1,18}$ & 573.0 & 1.27e-05  & 78    & 0.002 & - & - & blend \\
217405.0497 & 15$_{2,14}$$-$15$_{0,15}$ & 119.7 & 1.85e-06  & 62    & 0.005 & - & - & 2.7   \\
218893.9256 & 37$_{7,30}$$-$38$_{5,33}$ & 1135.4    & 1.27e-07  & 150   & 0.000 & - & - & blend \\
218894.1368 & 31$_{4,28}$$-$32$_{9,24}$ & 886.0 & 1.06e-08  & 130   & 0.000 & - & - & blend \\
220058.5907 & 19$_{1,18}$$-$19$_{1,19}$ & 553.8 & 1.53e-06  & 78    & 0.000 & - & - & blend \\
231114.2208 & 9$_{3,6}$$-$9$_{2,7}$ & 62.0  & 2.88e-05  & 38    & 0.064 & - & - & blend \\
231116.5868 & 12$_{8,5}$$-$11$_{8,4}$   & 421.5 & 2.42e-04  & 50    & 0.071 & - & - & 5.8   \\
231116.5868 & 12$_{8,4}$$-$11$_{8,3}$   & 421.5 & 2.42e-04  & 50    & 0.071 & - & - & 5.8   \\
231149.4163 & 12$_{11,1}$$-$11$_{11,0}$ & 344.6 & 6.98e-05  & 50    & 0.033 & - & - & blend \\
231150.2734 & 12$_{4,8}$$-$11$_{4,7}$   & 314.3 & 3.89e-04  & 50    & 0.226 & - & - & 8.6   \\
231212.6277 & 12$_{10,2}$$-$11$_{10,1}$ & 297.3 & 1.33e-04  & 50    & 0.086 & - & - & 10.1  \\
231212.6277 & 12$_{10,3}$$-$11$_{10,2}$ & 297.3 & 1.33e-04  & 50    & 0.086 & - & - & 10.1  \\
231255.9108 & 12$_{5,7}$$-$11$_{5,6}$   & 334.4 & 3.61e-04  & 50    & 0.184 & yes   & yes   & 8.8   \\
231268.3856 & 12$_{7,6}$$-$11$_{7,5}$   & 182.5 & 2.88e-04  & 50    & 0.389 & - & - & blend \\
231269.9018 & 12$_{6,7}$$-$11$_{6,6}$   & 153.4 & 3.28e-04  & 50    & 0.533 & - & - & 27.4  \\
231269.9031 & 12$_{6,6}$$-$11$_{6,5}$   & 153.4 & 3.28e-04  & 50    & 0.533 & - & - & blend \\
231270.2061 & 15$_{4,12}$$-$16$_{2,15}$ & 350.8 & 1.58e-07  & 62    & 0.000 & - & - & 27.4  \\
231329.6393 & 12$_{5,8}$$-$11$_{5,7}$   & 128.6 & 3.62e-04  & 50    & 0.689 & - & - & 27.4  \\
231329.7936 & 12$_{5,7}$$-$11$_{5,6}$   & 128.6 & 3.62e-04  & 50    & 0.689 & - & - & blend \\
231332.3518 & 19$_{19,1}$$-$18$_{17,2}$ & 1183.1    & 1.74e-07  & 78    & 0.000 & - & - & blend \\
231357.3496 & 12$_{3,9}$$-$11$_{3,8}$   & 299.1 & 4.11e-04  & 50    & 0.263 & yes   & yes   & 10.3  \\
231363.2836 & 12$_{5,7}$$-$11$_{5,6}$   & 128.5 & 3.62e-04  & 50    & 0.689 & - & - & 20.1  \\
231369.8292 & 12$_{5,8}$$-$11$_{5,7}$   & 128.5 & 3.62e-04  & 50    & 0.689 & yes   & - & 19.2  \\
231382.1065 & 12$_{5,8}$$-$11$_{5,7}$   & 333.6 & 3.62e-04  & 50    & 0.185 & yes   & - & 7.1   \\
231427.8171 & 4$_{4,0}$$-$5$_{3,2}$ & 45.4  & 1.85e-06  & 18    & 0.002 & - & - & 2.5   \\
231456.7437 & 12$_{4,9}$$-$11$_{4,8}$   & 108.4 & 3.90e-04  & 50    & 0.844 & yes   & yes   & 21.1  \\
231467.5036 & 12$_{4,8}$$-$11$_{4,7}$   & 108.4 & 3.90e-04  & 50    & 0.844 & yes   & - & 21.1  \\
231484.3739 & 12$_{4,8}$$-$11$_{4,7}$   & 108.3 & 3.90e-04  & 50    & 0.844 & yes   & yes   & 22.6  \\
232280.2669 & 32$_{4,28}$$-$31$_{5,26}$ & 733.0 & 7.02e-06  & 130   & 0.001 & - & - & blend \\
232290.3114 & 32$_{5,27}$$-$31$_{6,26}$ & 926.8 & 1.23e-08  & 130   & 0.000 & - & - & blend \\
232292.6027 & 12$_{4,9}$$-$11$_{4,8}$   & 311.8 & 3.92e-04  & 50    & 0.228 & yes   & yes   & 8.5   \\
234795.4555 & 12$_{2,10}$$-$11$_{2,9}$  & 81.9  & 4.45e-04  & 50    & 1.109 & - & - & blend \\
234823.4707 & 41$_{6,36}$$-$40$_{7,33}$ & 1254.2    & 1.75e-05  & 170   & 0.000 & - & - & blend \\
234825.8718 & 12$_{2,10}$$-$11$_{2,9}$  & 81.8  & 4.45e-04  & 50    & 1.110 & yes   & yes   & 22.2  \\
234902.0234 & 19$_{7,13}$$-$20$_{6,14}$ & 490.8 & 9.84e-06  & 78    & 0.003 & - & - & blend \\
234902.9702 & 12$_{2,10}$$-$11$_{2,9}$  & 287.2 & 4.47e-04  & 50    & 0.299 & - & - & 10.5  \\
234903.0341 & 19$_{7,12}$$-$20$_{6,15}$ & 490.8 & 9.84e-06  & 78    & 0.003 & - & - & blend \\
234904.3343 & 17$_{4,14}$$-$18$_{2,17}$ & 177.7 & 6.30e-08  & 70    & 0.000 & - & - & 10.5  \\
236065.5230 & 7$_{2,6}$$-$6$_{1,5}$ & 238.5 & 3.18e-06  & 30    & 0.002 & - & - & blend \\
236068.6485 & 28$_{9,20}$$-$29$_{8,21}$ & 557.9 & 1.12e-05  & 110   & 0.003 & - & - & 1.9   \\
236068.7156 & 28$_{9,19}$$-$29$_{8,22}$ & 557.9 & 1.12e-05  & 110   & 0.003 & - & - & blend \\
236071.8983 & 1$_{1,0}$$-$0$_{0,0}$ & 398.5 & 3.65e-05  & 6 & 0.001 & - & - & 2.4   \\
236527.7307 & 4$_{3,1}$$-$4$_{2,3}$ & 29.6  & 1.04e-07  & 18    & 0.000 & - & - & blend \\
237040.6540 & 35$_{4,31}$$-$35$_{3,32}$ & 623.6 & 6.83e-05  & 140   & 0.015 & - & - & 2.6   \\
\hline
    \end{tabular}
\tablefoot{Same for Table \ref{tab:tran_ch3oh} but for \ce{CH3CHO} with the corresponding figure as Figure \ref{fig:Fitting_ch3cho}.
}
    \label{tab:tran_ch3cho}
\end{table*}

\onecolumn
\LTcapwidth=\linewidth
\setlength{\tabcolsep}{4pt}
\begin{longtable}{cccccccccccc}
\caption{Transitions for the fitting - \ce{CH3OCHO}}\\
\hline\hline
Rest frequency
& Transition
& E$_{\rm up}$
& A$_{\rm ij}$
& upper state
& $\tau_{\rm 1 comp.}$
& Gau. fitting/stacked
& uv-space center
& S/N
\\
(MHz)
&	
& (K)
& (s$^{-1}$)
& degeneracy
& 
& 
&
&\\
\hline
\endfirsthead
\caption{continued.}\\
\hline\hline
Rest frequency
& Transition
& E$_{\rm up}$
& A$_{\rm ij}$
& upper state
& $\tau_{\rm 1 comp.}$
& Gau. fitting/stacked
& uv-space center
& S/N
\\
(MHz)
&	
& (K)
& (s$^{-1}$)
& degeneracy
& 
& 
&
&\\
\hline
\endhead
\hline
\endfoot
215420.8580 & 50$_{10,40}$$-$50$_{9,41}$    & 838.6 & 1.72e-05  & 200   & 0.004 & - & - & 1.6   \\
216109.7800 & 19$_{2,18}$$-$18$_{2,17}$ & 109.3 & 1.49e-04  & 78    & 1.843 & - & - & blend \\
216114.9599 & 29$_{9,20}$$-$29$_{8,21}$ & 312.0 & 1.51e-05  & 120   & 0.071 & - & - & blend \\
216115.5720 & 19$_{2,18}$$-$18$_{2,17}$ & 109.3 & 1.49e-04  & 78    & 1.843 & - & - & blend \\
216118.1536 & 33$_{5,28}$$-$33$_{5,29}$ & 542.5 & 5.25e-06  & 130   & 0.005 & - & - & blend \\
216282.5915 & 39$_{10,30}$$-$38$_{11,28}$   & 715.3 & 2.69e-06  & 160   & 0.001 & - & - & 1.7   \\
216518.4180 & 34$_{9,26}$$-$34$_{8,27}$ & 408.1 & 1.61e-05  & 140   & 0.045 & - & - & 3.6   \\
216588.6310 & 33$_{9,25}$$-$33$_{8,26}$ & 387.7 & 1.60e-05  & 130   & 0.048 & - & - & 2.6   \\
216958.8340 & 17$_{3,14}$$-$16$_{3,13}$ & 286.2 & 1.48e-04  & 70    & 0.484 & - & - & blend \\
216962.9890 & 20$_{0,20}$$-$19$_{1,19}$ & 111.5 & 2.45e-05  & 82    & 0.312 & - & - & blend \\
216964.1571 & 20$_{0,20}$$-$19$_{1,19}$ & 111.5 & 2.44e-05  & 82    & 0.311 & - & - & blend \\
216964.7650 & 20$_{1,20}$$-$19$_{1,19}$ & 111.5 & 1.53e-04  & 82    & 1.951 & - & - & blend \\
216965.9004 & 20$_{1,20}$$-$19$_{1,19}$ & 111.5 & 1.53e-04  & 82    & 1.952 & - & - & blend \\
216966.2462 & 20$_{0,20}$$-$19$_{0,19}$ & 111.5 & 1.53e-04  & 82    & 1.951 & - & - & blend \\
216967.4200 & 20$_{0,20}$$-$19$_{0,19}$ & 111.5 & 1.53e-04  & 82    & 1.952 & - & - & blend \\
216967.9947 & 20$_{1,20}$$-$19$_{0,19}$ & 111.5 & 2.45e-05  & 82    & 0.312 & - & - & blend \\
216969.1890 & 20$_{1,20}$$-$19$_{0,19}$ & 111.5 & 2.44e-05  & 82    & 0.311 & - & - & blend \\
217077.0790 & 30$_{4,26}$$-$30$_{4,27}$ & 291.5 & 5.12e-06  & 120   & 0.028 & - & - & 1.5   \\
217204.0112 & 34$_{7,28}$$-$34$_{5,29}$ & 572.5 & 4.96e-06  & 140   & 0.004 & - & - & 2.3   \\
217215.8470 & 32$_{9,24}$$-$32$_{8,25}$ & 367.8 & 1.59e-05  & 130   & 0.055 & - & - & 3.5   \\
217901.8163 & 38$_{10,28}$$-$38$_{9,29}$    & 692.5 & 1.45e-05  & 150   & 0.006 & - & - & 2.4   \\
218297.8900 & 17$_{3,14}$$-$16$_{3,13}$ & 99.7  & 1.51e-04  & 70    & 1.758 & - & - & blend \\
219153.3301 & 10$_{4,6}$$-$9$_{3,6}$    & 230.5 & 8.19e-06  & 42    & 0.023 & - & - & 5.1   \\
219154.5340 & 18$_{11,7}$$-$17$_{11,6}$ & 369.0 & 9.94e-05  & 74    & 0.190 & - & - & blend \\
219566.2450 & 18$_{15,3}$$-$17$_{15,2}$ & 438.1 & 4.89e-05  & 74    & 0.058 & - & - & blend \\
219566.2450 & 18$_{15,4}$$-$17$_{15,3}$ & 438.1 & 4.89e-05  & 74    & 0.058 & - & - & blend \\
219568.4800 & 18$_{14,5}$$-$17$_{14,4}$ & 418.7 & 6.32e-05  & 74    & 0.085 & - & - & 6.1   \\
219568.4800 & 18$_{14,4}$$-$17$_{14,3}$ & 418.7 & 6.32e-05  & 74    & 0.085 & - & - & 6.1   \\
219571.1980 & 18$_{16,3}$$-$17$_{16,2}$ & 458.9 & 3.36e-05  & 74    & 0.034 & - & - & blend \\
219571.1980 & 18$_{16,2}$$-$17$_{16,1}$ & 458.9 & 3.36e-05  & 74    & 0.034 & - & - & blend \\
220523.4125 & 36$_{9,27}$$-$35$_{10,26}$    & 636.4 & 1.44e-06  & 150   & 0.001 & - & - & blend \\
220525.2148 & 10$_{4,6}$$-$9$_{3,7}$    & 230.8 & 1.03e-05  & 42    & 0.029 & - & - & 2.4   \\
220671.4458 & 29$_{9,20}$$-$29$_{8,21}$ & 497.9 & 1.23e-05  & 120   & 0.015 & - & - & 2.3   \\
221047.7910 & 18$_{14,4}$$-$17$_{14,3}$ & 230.9 & 6.45e-05  & 74    & 0.314 & - & - & blend \\
221047.7910 & 18$_{14,5}$$-$17$_{14,4}$ & 230.9 & 6.45e-05  & 74    & 0.314 & - & - & blend \\
221049.9900 & 18$_{14,4}$$-$17$_{14,3}$ & 230.9 & 6.45e-05  & 74    & 0.314 & - & - & blend \\
221064.4128 & 7$_{3,5}$$-$6$_{0,6}$ & 22.5  & 3.18e-07  & 30    & 0.003 & - & - & blend \\
221066.9330 & 18$_{14,5}$$-$17$_{14,4}$ & 230.9 & 6.45e-05  & 74    & 0.314 & - & - & blend \\
221075.9810 & 29$_{9,21}$$-$29$_{8,22}$ & 312.0 & 1.61e-05  & 120   & 0.073 & - & - & blend \\
221082.0006 & 10$_{5,5}$$-$10$_{3,8}$   & 49.1  & 3.43e-07  & 42    & 0.003 & - & - & blend \\
221086.1780 & 29$_{9,21}$$-$29$_{8,22}$ & 312.0 & 1.61e-05  & 120   & 0.073 & - & - & 3.2   \\
231045.9845 & 12$_{4,9}$$-$11$_{3,8}$   & 56.8  & 1.03e-05  & 50    & 0.103 & - & - & 3.9   \\
231047.4058 & 36$_{25,12}$$-$37$_{24,13}$   & 995.6 & 9.42e-07  & 150   & 0.000 & - & - & blend \\
231047.5117 & 36$_{25,11}$$-$37$_{24,14}$   & 995.6 & 9.42e-07  & 150   & 0.000 & - & - & blend \\
233669.9773 & 38$_{9,30}$$-$37$_{10,28}$    & 680.5 & 1.59e-06  & 150   & 0.001 & - & - & blend \\
233670.9800 & 19$_{12,8}$$-$18$_{12,7}$ & 207.6 & 1.16e-04  & 78    & 0.626 & yes   & yes   & 10.7  \\
233845.2330 & 19$_{11,8}$$-$18$_{11,7}$ & 192.4 & 1.29e-04  & 78    & 0.770 & yes   & yes   & 12.0  \\
233854.2860 & 19$_{11,8}$$-$18$_{11,7}$ & 192.4 & 1.29e-04  & 78    & 0.770 & - & - & 17.2  \\
233854.2860 & 19$_{11,9}$$-$18$_{11,8}$ & 192.4 & 1.29e-04  & 78    & 0.770 & - & - & 17.2  \\
233867.1930 & 19$_{11,9}$$-$18$_{11,8}$ & 192.4 & 1.29e-04  & 78    & 0.770 & yes   & yes   & 12.4  \\
233959.0100 & 16$_{9,8}$$-$16$_{8,9}$   & 134.0 & 1.39e-05  & 66    & 0.105 & - & - & 4.4   \\
234674.9700 & 19$_{9,11}$$-$19$_{8,12}$ & 353.1 & 1.59e-05  & 78    & 0.031 & - & - & 1.3   \\
234735.6500 & 13$_{9,5}$$-$13$_{8,6}$   & 107.4 & 1.13e-05  & 54    & 0.084 & - & - & blend \\
234737.4750 & 9$_{5,5}$$-$8$_{4,4}$ & 43.2  & 1.70e-05  & 38    & 0.138 & - & - & 7.8   \\
234739.1110 & 20$_{2,18}$$-$19$_{3,17}$ & 128.0 & 1.98e-05  & 82    & 0.193 & - & - & blend \\
234915.0220 & 36$_{7,30}$$-$36$_{5,31}$ & 430.1 & 6.93e-06  & 150   & 0.015 & - & - & blend \\
234916.8050 & 9$_{5,4}$$-$8$_{4,5}$ & 43.2  & 1.70e-05  & 38    & 0.138 & - & - & 4.9   \\
235122.0070 & 10$_{9,2}$$-$10$_{8,3}$   & 86.2  & 6.43e-06  & 42    & 0.043 & - & - & 3.0   \\
235144.4940 & 10$_{9,1}$$-$10$_{8,2}$   & 86.2  & 6.42e-06  & 42    & 0.043 & - & - & 3.6   \\
235145.0245 & 44$_{8,36}$$-$44$_{8,37}$ & 644.6 & 7.04e-06  & 180   & 0.004 & - & - & blend \\
236108.2250 & 21$_{1,20}$$-$20$_{2,19}$ & 318.0 & 2.60e-05  & 86    & 0.071 & - & - & 4.9   \\
236465.2486 & 41$_{7,34}$$-$41$_{6,35}$ & 739.5 & 1.85e-05  & 170   & 0.005 & - & - & 1.1   \\
236492.4650 & 34$_{6,29}$$-$34$_{5,30}$ & 377.6 & 1.53e-05  & 140   & 0.045 & - & - & blend \\
236683.7060 & 17$_{9,9}$$-$17$_{8,10}$  & 331.0 & 1.49e-05  & 70    & 0.030 & - & - & 2.1   \\
236759.6870 & 19$_{5,15}$$-$18$_{5,14}$ & 129.6 & 1.86e-04  & 78    & 1.679 & yes   & yes   & 20.6  \\
237056.9149 & 41$_{10,32}$$-$40$_{11,29}$   & 763.9 & 1.38e-06  & 170   & 0.000 & - & - & blend \\
237057.1470 & 9$_{5,4}$$-$8$_{4,4}$ & 230.6 & 1.69e-05  & 38    & 0.037 & - & - & 2.1   \\
237879.9308 & 17$_{6,12}$$-$17$_{4,13}$ & 114.6 & 1.02e-06  & 70    & 0.009 & - & - & 2.3  
    \label{tab:tran_ch3ocho}
\end{longtable}
\tablefoot{
    Same for Table \ref{tab:tran_ch3oh} but for \ce{CH3OCHO} with the corresponding figure as Figure \ref{fig:Fitting_ch3ocho}.
}
\twocolumn

\section{\ce{CH3OCH3}}
\label{app:CH3OCH3}
\begin{figure*}
\centering
\includegraphics[width=1\textwidth]{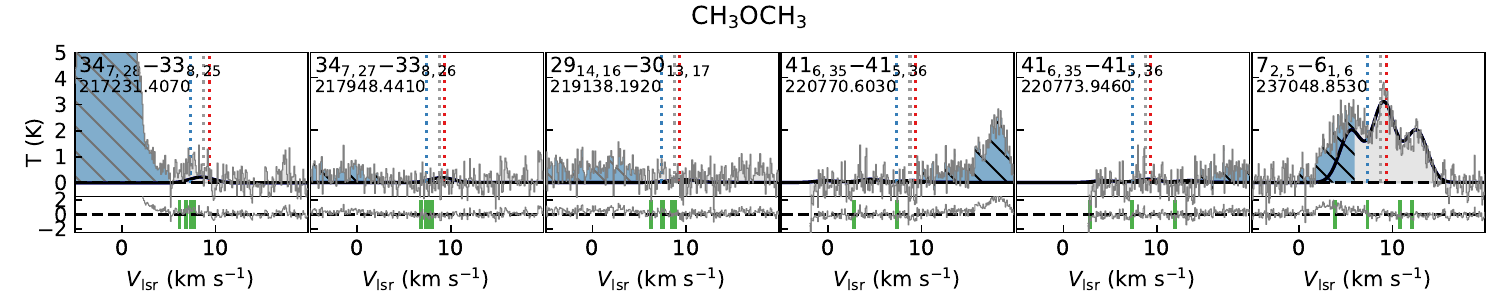}
\caption{Same as Figure \ref{fig:Fitting_ch2oh2} but for \ce{CH3OCH3}.
}
\label{fig:Fitting_ch3och3}
\end{figure*}

\ce{CH3OCH3} is detected in the narrowband high-resolution data in only a few transitions without contamination from other lines (Figure \ref{fig:Fitting_ch3och3}). The only detection with high S/N ratio, $7_{2,5}-6_{1,6}$, is blended by other transitions so that we cannot analyse the kinematics like in Figures \ref{fig:stack} and \ref{fig:uv_cen}. One-component LTE fitting is conducted, yielding a source with $T_{\rm ex}=173$ K, $V_{\rm lsr}=9.1$ km s$^{-1}$, and FWHM$=2.71$ km s$^{-1}$. The source size and column density is not well constrained given all optically thin ($\tau<0.2$) line emission.
\ce{CH3OCH3} and \ce{CH3OCHO} are considered to be chemically linked (for example \citealt{co20}). The one-component models do not provide the same temperature and $V_{\rm lsr}$, but this could simply be due to the very 
limited number of transitions available in the case of  \ce{CH3OCH3}.

\end{document}